\documentclass[aps,prd,twocolumn,amsmath,amssymb,floatfix,nofootinbib,superscriptaddress]{revtex4-1}

\usepackage{mathtools}
\usepackage{graphicx}
\usepackage{dcolumn}
\usepackage{bm}
\usepackage[usenames, dvipsnames]{color}
\usepackage[normalem]{ulem}
\usepackage[dvipsnames]{xcolor}
\usepackage[utf8]{inputenc}
\usepackage[colorlinks=true,citecolor=blue,urlcolor=blue]{hyperref}
\usepackage{aas_macros}

\newcommand{\be}{\begin{align}}

\newcommand{\ee}{\end{align}}

\usepackage{physics}
\usepackage{amsmath}
\usepackage{tikz}
\usepackage{mathdots}
\usepackage{yhmath}
\usepackage{cancel}
\usepackage{color}
\usepackage{siunitx}
\usepackage{array}
\usepackage{multirow}
\usepackage{amssymb}
\usepackage{gensymb}
\usepackage{tabularx}
\usepackage{extarrows}
\usepackage{booktabs}
\usetikzlibrary{fadings}
\usetikzlibrary{patterns}
\usetikzlibrary{shadows.blur}
\usetikzlibrary{shapes}
\usepackage{mathtools}
\usepackage{tikz}

\DeclareMathOperator{\sinc}{sinc}

\newcommand{\smu}{Department of Physics,
Southern Methodist University, 3215 Daniel Ave, Dallas, TX 75275, USA}
\newcommand{\utaustin}{Center for Gravitational Physics, University of Texas at Austin, 2515 Speedway, C1600, Austin, TX 78712, USA}

\begin{document}

\title{Gravitational Wave Timing Array}
\author{Mar\'ia~Jos\'e~Bustamante-Rosell}
\affiliation{\utaustin}
\author{Joel~Meyers}
\affiliation{\smu}
\author{Noah~Pearson}
\affiliation{\smu}
\author{Cynthia~Trendafilova}
\affiliation{\smu}
\author{Aaron~Zimmerman}
\affiliation{\utaustin}

\date{\today}

\begin{abstract}
We describe the design of a gravitational wave timing array, a novel scheme that can be used to search for low-frequency gravitational waves by monitoring continuous gravitational waves at higher frequencies.
We show that observations of gravitational waves produced by Galactic binaries using a space-based detector like LISA provide sensitivity in the nanohertz to microhertz band. 
While the expected sensitivity is several of magnitude worse than what can be achieved by pulsar timing arrays, it supplements other recent proposals for gravitational wave searches in the microhertz regime. 
This regime is below the frequencies to which LISA is directly sensitive, and above the frequency range generally targeted by pulsar timing array searches.
The low-frequency extension of sensitivity does not require any experimental design change to space-based gravitational wave detectors, and can be achieved with the data products that would already be collected by them.
\end{abstract}

\maketitle

\section{Introduction}

Current and future observatories probe gravitational waves in several regimes of the frequency spectrum.
Ground-based gravitational wave detectors, such as the Advanced LIGO~\cite{TheLIGOScientific:2014jea}, Advanced Virgo~\cite{TheVirgo:2014hva}, and KAGRA~\cite{Akutsu:2020his} detectors, cover the $\sim 20-2000$ Hz range by measuring induced gravitational wave strains of order $10^{-22}$ in their kilometer-scale Fabry-Pérot interferometers.
In the near future, the Laser Interferometer Space Antenna (LISA) will probe the millihertz regime, measuring gravitational waves using million kilometer-scale arms~\cite{Audley:2017drz}.
At nanohertz frequencies, 
pulsar timing arrays indirectly measure gravitational waves 
by monitoring an array of pulsars which serve as standard clocks~\cite{Verbiest:2021kmt,
NANOGrav:2020qll,Kerr:2020qdo,Babak:2015lua,Hobbs:2009yy}.
Strongly lensed, repeating fast radio bursts~\cite{Petroff_2019} can also be utilized to detect gravitational waves in a similar manner to pulsar timing arrays, by monitoring changes to the arrival times of the lensed burst images~\cite{PhysRevD.103.063017}.
Astrometric observations of distant objects can be used to search for photon deflection caused by gravitational waves in the nanohertz regime, along with an integrated constraint on a background of lower frequency gravitational waves~\cite{1990NCimB.105.1141B,Pyne_1996,Book:2010pf}.
A key goal of current and future cosmic microwave background surveys is the search for primordial gravitational waves with frequencies in the attohertz range through measurements of $B$-mode polarization~\cite{Kamionkowski:1996zd,Seljak:1996gy}.

These techniques, together with current and future observatories, cover a wide swath of the gravitational wave spectrum.
A gap remains between the lowest frequencies accessible on the ground, $\sim1$~Hz, and the highest frequencies accessible to LISA. 
Several proposed observatories plan to cover the gap between ground-based interferometers and LISA, including the DECi-hertz Interferometer Gravitational wave Observatory (DECIGO)~\cite{2017CEAS....9..371M,2020arXiv200613545K} and the Big Bang Observer (BBO)~\cite{Crowder:2005nr}, and that regime may also be accessible with atom interferometry~\cite{Canuel:2017rrp,Canuel:2019abg}.
Meanwhile, the sensitivity of LISA is limited to frequencies $\gtrsim 10^{-5}$\textendash{}$10^{-4}$~Hz by acceleration performance~\cite{Larson2005,Babak:2021mhe},
while pulsar timing arrays usually focus on the frequency regime $< 10^{-6}$~Hz, limited by the cadence with which pulsars in the network are observed.
The sensitivity of pulsar timing arrays can be extended to higher frequencies by timing pulsars at a higher cadence~\cite{Perera:2018pts}.
A recent study demonstrates that staggered, lower-cadence observations of many pulsars can also extend the sensitivity of pulsar timing arrays to higher frequencies, $\gtrsim 10^{-6}$~ Hz~\cite{Wang:2020hfh}.
This microhertz regime has been targeted by at least one proposal~\cite{Sesana:2019vho} but remains unlikely to be covered by direct gravitational wave searches in the next few decades.
Additionally, an interesting recent proposal showed that high-cadence astrometric measurements obtained from photometric surveys can be used to search for gravitational waves with frequencies ranging from nanohertz to microhertz~\cite{Wang:2020pmf}. 
Another recent proposal suggests that precise tracking of orbital dynamics can be used to detect gravitational waves over a wide range of frequencies, $~10^{-8}$\textendash{}$10^{-4}$~Hz \cite{Blas:2021mqw,Blas:2021mpc}.
Yet another recent proposal suggests using $\sim10$~km diameter asteroids as natural test masses with low acceleration noise to search for gravitational waves with frequencies in the $10^{-7}$\textendash{}$10^{-5}$~Hz range \cite{Fedderke:2021kuy}.

In this paper, we describe a new method that can be used to search for gravitational waves in the microhertz band.
Gravitational waves emitted from Galactic binaries observed with space-based interferometers like LISA act as stable oscillators that can be monitored to construct a gravitational wave timing array.
Within the LISA band, the most common source of gravitational waves will be the millions of white dwarf binaries present in the Milky Way.
About ten thousand of them should be well resolved by the LISA mission within its planned four-year lifetime. 
Additionally, it is possible that LISA will detect other classes of binaries emitting in the mHz regime, such as mixed binaries of white dwarfs and neutron stars or black holes~\cite{Korol:2017qcx,Breivik:2019lmt}.
Most of these Galactic binaries will emit nearly monochromatic gravitational waves during LISA's lifetime, with the evolution of the binaries dominated by gravitational wave radiation.

\begin{figure}[tb]
    \centering
    \includegraphics[width=\columnwidth]{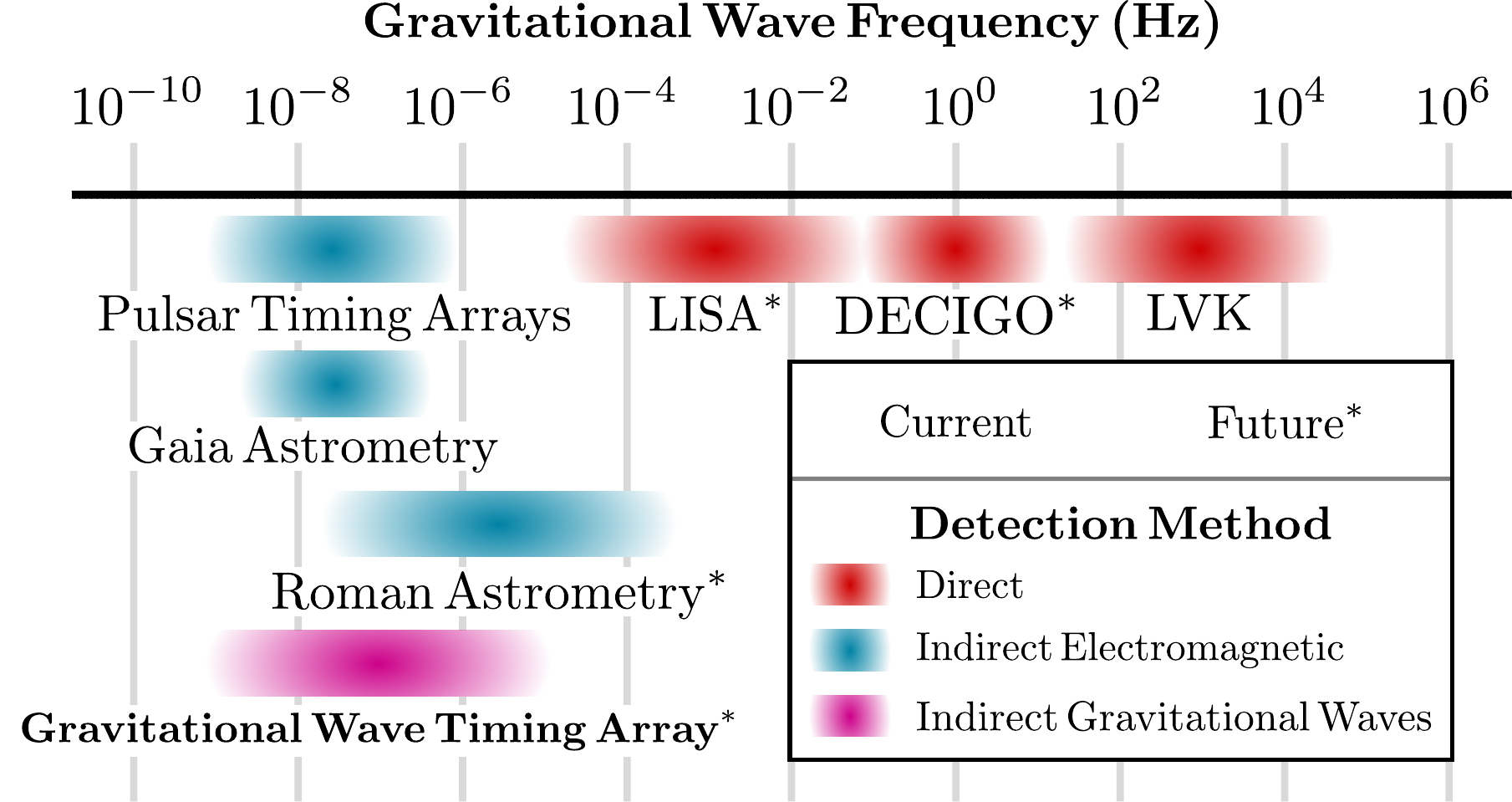}
    \caption{
    Targeted frequency range of several existing and proposed gravitational wave detection methods.
    Direct detection methods (red) include the existing ground-based detectors LIGO~\cite{TheLIGOScientific:2014jea}, Virgo~\cite{TheVirgo:2014hva}, and Kagra~\cite{Akutsu:2020his} (LVK), the space-based LISA, and proposed DECIGO~\cite{2020arXiv200613545K} detectors.
    Indirect methods (blue) include existing pulsar timing arrays~\cite{Verbiest:2021kmt}, astrometry with Gaia observations~\cite{2018CQGra..35d5005K}, and proposed high-cadence astrometry with the future Nancy Grace Roman Space Telescope~\cite{Wang:2020pmf,2019JATIS...5d4005W}.
    The gravitational wave timing array proposed here (purple) would use LISA observations to detect gravitational waves indirectly.
    }
    \label{fig:frange}
\end{figure}

A gravitational wave background, either stochastic or coherent, imprints correlated phase modulations on the gravitational waves from the binaries, similar to the timing delays measured in pulsar timing arrays.
A gravitational wave timing array like the one described here benefits from the large number of sources and continuous monitoring of the whole sky, and it provides sensitivity to gravitational waves in a frequency range $10^{-9}$--$10^{-5}$~Hz.
The frequency coverage of such a gravitational wave timing array is illustrated in Fig.~\ref{fig:frange}, along with several existing experiments and future proposals for gravitational wave detection.
Constructing such an array with the observations from a space-based interferometer like LISA requires no experimental design changes nor a specialized observing campaign.
Gravitational waves are not subject to any plasma dispersion effects in the interstellar medium or radio interference at Earth that can complicate gravitational wave searches with pulsar timing arrays~\cite{Levin_2015}.
On the other hand, a gravitational wave timing array relies on intrinsically weak gravitational waves as the primary signal to be monitored, and this leads to lower sensitivity than direct LISA measurements and pulsar timing arrays in the regimes where those strategies are sensitive.

This paper is organized as follows. 
In Sec.~\ref{sec:Mod} we discuss the phase modulation induced on a gravitational wave propagating in a flat spacetime which is perturbed by an additional background gravitational wave. 
In Sec.~\ref{sec:Sensitivity} we calculate the sensitivity of a gravitational wave timing array using multiple methods.
Section~\ref{sec:TimeDomainSensitivity} presents a timing estimate approach leveraging existing results for pulsar timing arrays~\cite{Moore:2014eua}, Sec.~\ref{subsec:MatchedFilter} provides a matched filter sensitivity estimate,  and Sec.~\ref{sec:Fisher} describes our most complete frequency-domain Fisher estimates.
Section~\ref{sec:mock} uses a mock Galaxy catalog and the results from Sec.~\ref{sec:Sensitivity} to get realistic estimates of the sensitivity of a gravitational wave timing array constructed from LISA observations over a nominal four-year mission. 
Lastly, Sec.~\ref{sec:discussion} discusses the viability of such an array, the limitations of our approximations, and future prospects.

\section{Modulation of a carrier wave by a background wave}
\label{sec:Mod}

\begin{figure}[tb]
    \centering
    \includegraphics[width=0.8\columnwidth]{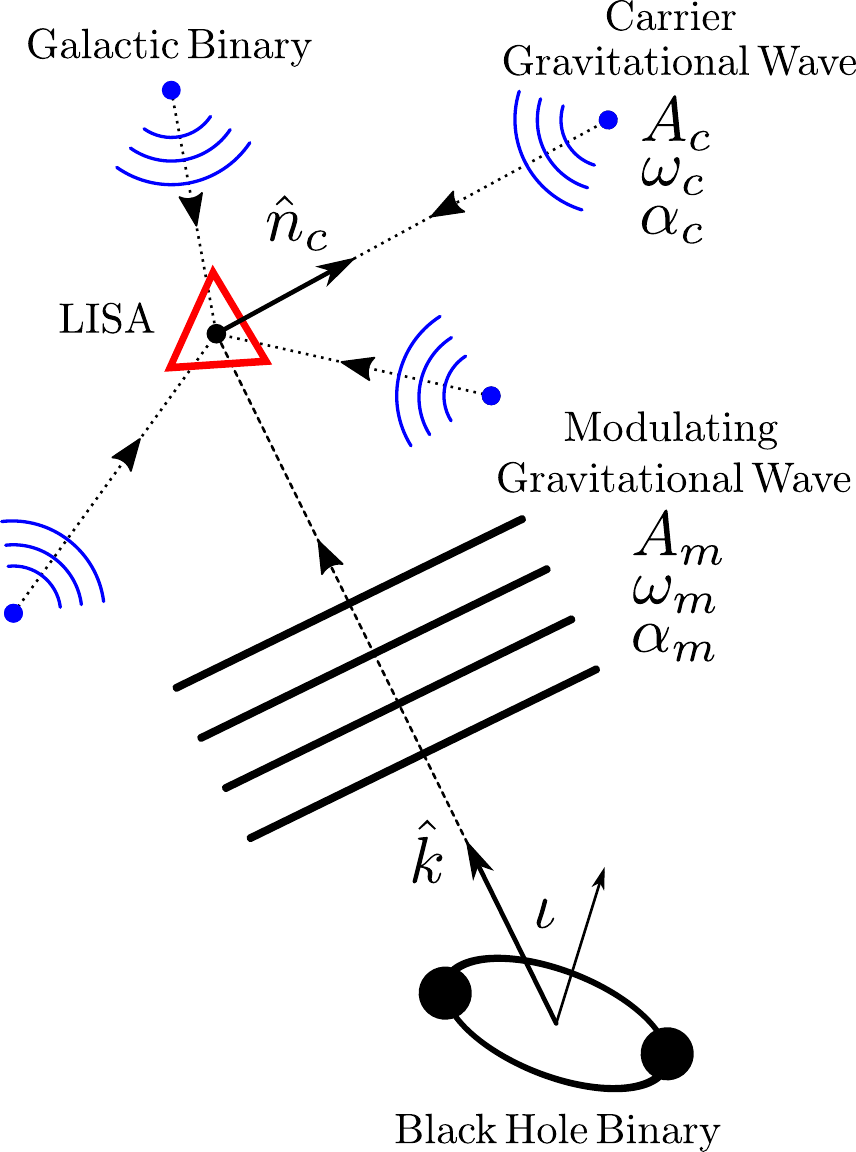}
    \caption{Illustration of the proposed gravitational wave timing array. An array of Galactic binaries within the Milky Way produce gravitational waves whose frequencies are modulated by a longer-wavelength gravitational wave, exemplified here by one produced by a black hole binary. Unit vector $\hat{n}_c$ points from the observer (such as LISA) to each galactic binary, and $\hat{k}$ denotes the propagation direction of the modulating gravitational wave. The inclination of the binary black hole is $\iota$, such that the binary is face-on for $\iota = 0$. $A_c,\,A_m,\,\omega_c,\,\omega_m,\,\alpha_c,\,\alpha_m,\,$ are the amplitudes, angular frequencies, and phases of the carrier and modulating gravitational waves, respectively.}
    \label{fig:geo_cartoon}
\end{figure}

We consider an approximately monochromatic gravitational wave with frequency $\omega_c$, which we refer to as the carrier wave.
This wave propagates in the presence of a lower frequency, modulating gravitational plane wave, with frequency $\omega_m\ll \omega_c$.
To leading order in the modulating wave amplitude, the observed phase evolution of the carrier wave is
\begin{align}
    \frac{d\varphi}{dt} \approx \omega_c \left[1 - z(t, \hat k)\right] \,,
\end{align}
where $\varphi$ is the phase of the carrier, and $z$ is a redshift induced by the modulating gravitational wave.
It is given by~\cite{Book:2010pf,Maggiore:2018sht}
\begin{align}
z(t,\hat k) & = 	
\frac{\hat n^i \hat n^j}{2 (1+\hat k \cdot \hat n ) }  \left[ h_{ij}(t,\hat k)-h_{ij}(t_c,\hat k) \right] \,,
	\label{eq:redshift}
\end{align}
where $\hat k$ is the direction of propagation of the modulating wave and $\hat n$ is the unit vector pointing from the observer towards the source of the carrier waves, as shown in Fig.~\ref{fig:geo_cartoon}. 

The two terms $h_{ij}(t,\hat k)$ and $h_{ij}(t_c,\hat k)$ are the metric perturbations from the modulating wave at the observer and at the source of the carrier wave, respectively.
In the context of pulsar timing arrays, the first term is called the Earth term, and the second the pulsar term.
The time $t_c = t - d(1 + \hat n \cdot \hat k)/c $ incorporates the relative propagation times of the carrier wave and modulating waves, where $d$ is the distance between the observer and the source of the carrier waves.

When considering the timing residuals of pulsars, the redshift of Eq.~\eqref{eq:redshift} is derived by considering the propagation of a null geodesic between the source of the pulse and the observer~\cite{Book:2010pf,Anholm_2009}.
The same result is arrived at in our case by noting that for $\omega_m \ll \omega_c$, the carrier wave is in the geometric optics limit, so that the wavefronts propagate along geodesic null rays, as discussed further in Appendix~\ref{sec:GeometricOptics}.
The phase of the carrier wave is then given by 
\begin{align}
	\varphi (t,\hat k )&\approx \omega_c t - \omega_c \int_0^t z (t',\hat k ) \dd t' \,.
\end{align}
When searching for a stochastic background of gravitational waves with pulsar timing arrays, the pulsar term in Eq.~\eqref{eq:redshift} is a nuisance parameter that contributes to the timing noise for each pulsar.
It can often be neglected when estimating the sensitivity of the array.
Since we consider a binary source of modulating waves, we find that this pulsar term potentially contributes to the detected signal.
Whether we can neglect this term, which we call the carrier term, will thus depend on the evolution of the modulating wave.

\subsection{Source of modulating waves}

We assume that the source of the modulating wave is a supermassive black hole binary, which produces a transverse-traceless gravitational wave
\begin{align}
\label{eq:hTT}
	h_{ij} ( t,\hat k )= A_m^+H^+_{ij} ( \hat k ) \cos\Phi (t) +A_m^\times H^\times_{ij} ( \hat k ) \sin \Phi (t)\,,
\end{align}
where $\Phi(t)$ is the phase of the modulating wave.
The binary has an inclination angle $\iota$ relative to the line of sight, so that the amplitudes of the plus- and cross-polarized waves are related to a characteristic amplitude $A_m$ through 
\begin{align}
\label{eq:PolarizationAmps}
    A_m^+ & = A_m\frac{1+\cos^2 \iota}{2}\,, 
    & A_m^\times = A_m\cos \iota \,.
\end{align}
For later reference, we define two preferred transverse polarization tensors $\epsilon^+_{ij} ( \hat k )$ and $\epsilon^\times_{ij} ( \hat k )$, and a polarization angle $\psi$.
Then
\begin{align}
    H^+_{ij} & = \cos 2\psi \, \epsilon^+_{ij} 
    + \sin 2\psi \, \epsilon^\times_{ij}\,, \\
    H^\times_{ij} & = -\sin 2\psi\, \epsilon^+_{ij} + \cos 2\psi \, \epsilon^\times_{ij} \,.
\end{align}

Assuming that the modulating binary evolves only due to the emission of gravitational waves, its frequency evolution is given by
\begin{align}
\label{eq:FrequencyEvolution}
\dot \omega_m & =\frac{12}{5}   \left(\frac{G \mathcal M}{c^3}\right)^{5/3} \omega_m ^{11/3}
\end{align}
where $\mathcal M = (m_1 m_2)^{3/5}/(m_1+m_2)^{1/5}$ is the chirp mass of the binary, $m_1$ is the mass of the primary, and $m_2$ that of the secondary.
The modulating wave is approximately monochromatic with frequency $\omega_m$ over the timescale of the observation, which is of the order of years (the lifetime of the LISA mission).
Equations~\eqref{eq:redshift}--\eqref{eq:FrequencyEvolution} show that the phase of the carrier wave is modulated sinusoidally by the background wave, potentially at two distinct frequencies, one associated with the local term and one with the carrier term.
The rate of change of $\omega_m$ is a steep function of $\omega_m$, and so at a fixed chirp mass we see that there are two cases.
Either the frequency evolves very slowly, even on timescales sufficiently long for light to propagate across the Galaxy ($\sim 10$~kyr), or else the frequency evolves rapidly over such timescales.
In the first case, we cannot neglect the carrier term, while in the second we can.
See Refs.~\cite{Sesana:2010ac,Babak:2015lua} for similar discussion of the monochromatic limit, as well as \cite{Maggiore:2018sht} (Ch.~23.2).

\subsection{Slow evolution of modulating source}

First consider the case where we neglect the evolution of the modulating binary.
Then both the local (Earth) term and the carrier (pulsar) term of the modulation have the same frequency, but a relative phase difference due to propagation effects.
The various sinusoidal terms combine to give a redshift
\begin{align}
	z & = A_m F \sin\gamma \cos(\omega_m t - \delta)\,,
	\label{eq:genterm}
\end{align}
where
\begin{align}
	F &=\frac{\sqrt{(A_m^+H_{nn}^+)^2+(A_m^\times H_{nn}^\times)^2 }}{A_m(1+\hat k \cdot \hat n)} \,,
	\label{eq:genamp}
	\\
	\gamma & = \frac{\omega _m d}{2 c}(1 + \hat k \cdot \hat n) \,,
	\label{eq:genmodamp}
	\\
	\delta & = \alpha_m + \gamma + \beta - \frac{\pi}{2}  \,, 
	\label{eq:genphase}
	\\
	\tan \beta & = \frac{A_m^\times H_{nn}^\times}{A_m^+ H_{nn}^+} \,,
\end{align}
and where $H_{nn}^+ = H^+_{ij}\hat n^i \hat n^j$ is the plus-polarized component of the wave projected onto $\hat n$, and similarly for $H_{nn}^\times$.
We have introduced a phase $\alpha_m$, which is a reference phase for the modulating binary, a phase $\beta$ which varies with $\hat n$ at fixed $\hat k$, and a phase $\gamma$ which also depends on the distance to the source of the carrier.

Overall this gives a modulated carrier wave phase
\begin{align}
	\varphi(t,\hat k)
	& \approx \omega_c t  - \alpha_c 
	- \frac{A_m F \omega_c}{\omega_m} \sin \gamma \sin(\omega_m t - \delta) \,,
	\label{eq:phase}
\end{align}
where we have introduced another reference phase $\alpha_c$ to absorb the integration constant.
Equation~\eqref{eq:phase} shows that the modulation of the phase is directly proportional to number of cycles of the carrier that occur per half a cycle of the modulating wave.

\begin{figure}[tb]
	\centering
	\includegraphics[width=\columnwidth]{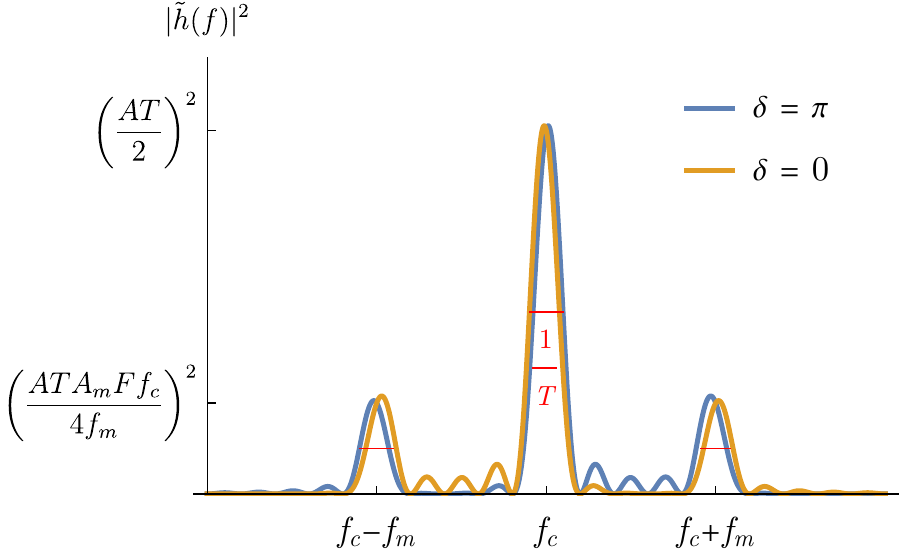}
	\caption{Schematic representation of the signal in the frequency domain. The full width at half maximum of each peak is $\sim 1/T$ (highlighted with a red line), where $T$ is the total observation time.
	}
	\label{fig:freq_diag}
\end{figure}

We mostly work in the frequency domain, where the carrier signal is a sharp peak at a frequency $f = f_c$, and another peak mirrored around zero at $f = -f_c$ as required for a real signal.
The modulation produces a sideband peak
to each side of these carrier peaks, separated by $f_m$ as shown in Fig.~\ref{fig:freq_diag}. 
The redshift phase $\delta$ introduces an asymmetry in the two sideband peaks around each carrier peak.
Since we deal with a finite observation time $T$, these peaks are sinc functions rather than delta functions, with finite widths in frequency space of order $1/T$.

In the limit that the modulating binary does not evolve, both the phase and amplitude of the modulating signal depend on $d$, through $\gamma$.
Since the distance $d$ to any Galactic binary is unknown to the precision of a wavelength of the modulating wave, $\gamma$ is a nuisance parameter that degrades our ability to detect the influence of a modulating wave on the carrier waves coming from a network of Galactic binaries, all at different unknown distances.

However, if the frequency of the modulating wave evolves over typical timescales $\sim d/c$, then the local term and the carrier term each produce distinct sidebands around the carrier.
When the frequency of the modulating wave evolves fast enough that the sidebands produced by the local term and the carrier term are well-separated in frequency space, the local term will provide a pair of sidebands around each carrier peak whose amplitude does not depend on $d/c$.
We turn to this case next.

\subsection{Fast evolution of modulating source}
\label{sec:FastEvolve}

For the range of frequencies that we are interested in, we can determine the parameter space where the modulating frequency evolves enough to separate the local and carrier sidebands for each Galactic binary. 
The local term would then be common to all the carrier waves, up to a phase and geometric factors that do not depend upon the distance $d$ to any binary. 
Meanwhile the carrier terms differ in frequency and amplitude among the Galactic binaries, and can be neglected as extra noise.

In order to treat the local and carrier sidebands as well separated, we require that the change in the modulation frequency over the propagation time of the carrier wave be larger than the full width at half maximum (FWHM) of the sideband peak in the frequency domain,
\begin{align}
\mathrm{FWHM} \left[ \sinc(\pi f_m T) \right] < \dot f_m d/c \,.
\end{align}
Combining with Eq.~\eqref{eq:FrequencyEvolution}, we see that the modulating frequencies for which the local and carrier terms are well separated satisfy
\begin{align}
	f_m > 0.46 \, \mu{\rm Hz}
	\left(\frac{ {\rm 0.1 \, kpc}}{d}\right)^{3/11}
	\left(\frac{10^6M_\odot}{\mathcal M}\right)^{5/11}
	\left(\frac{ {4 \, \rm {yr}}}{T}\right)^{3/11}\,.
\end{align}
We have chosen a small fiducial chirp mass and distance to display the largest frequency for which the carrier (pulsar) terms can be neglected.

For modulating frequencies higher than this, 
the phase of the Galactic binary wave in the presence of a modulating wave simplifies to
\begin{align}
	\varphi(t,\hat k)
	& \approx \omega_c t  - \alpha_c 
	- \frac{A_m F \omega_c}{2\omega_m} \sin(\omega_m t -\alpha_m - \beta) \,.
	\label{eq:phaseNoPulsar}
\end{align}
Comparing to the case where we include the carrier term, we recover Eq.~\eqref{eq:phaseNoPulsar} from Eq.~\eqref{eq:phase} by setting $\gamma = \pi/2$ and adding a factor of $1/2$ to the modulating part.
Further, if we fix the location of the modulating binary along with its inclination and polarization angles, $F$ and $\beta$ depend only on the sky location of the white dwarf binary and not its distance from Earth.
In practice, the sky position of a white dwarf binary detected by LISA can be determined to a few square degrees (see Sec.~\ref{sec:mock}), and so we treat $F$ and $\beta$ as approximately known in what follows.

For later use, we define a convenient coordinate system in which to evaluate the geometric quantities $F$ and $\beta$. 
Let the modulating wave propagate in the $z$ direction, and set the polarization tensors to be $\epsilon^+_{ij}= \hat e^x_i \hat e^x_j - \hat e^y_i \hat e^y_j$ and $\epsilon^+_{ij}= \hat e^x_i \hat e^y_j + \hat e^y_i \hat e^x_j$, where $\hat e^x$ and $\hat e^y$ are unit vectors in the $x$ and $y$ directions. 
In this basis, the redshift will be in terms of $\theta$ and $\phi$, the colatitude and longitude of the carrier source,
\begin{align} 
\label{eq:general_geometry_F}
    F & =\sin ^2 \frac{\theta}{2}\bigg(\cos^2(2 \phi - 2 \psi)\sin^4 \iota  + 4 \cos^2 \iota \bigg)^{1/2} \,, \\
	\delta & =  \alpha_m+ \tan^{-1} \left[\frac{2\cos \iota \tan(2\phi - 2 \psi)}{1+ \cos^2 \iota} \right]\,.
\end{align}

We can further specialize to the two extremes of the inclination.
When the binary is face-on, $\iota = 0$, the waves are circularly polarized, and
\begin{align}
    F & = 2 \sin ^2 \frac{\theta}{2}\,, & \beta & = 2 (\phi -\psi )\,. 
\end{align}
When the binary is edge-on, the modulating wave is linearly polarized, and
\begin{align}
    F & =\sin ^2 \frac{\theta}{2} \cos (2 \psi -2 \phi )\,, & \beta & = 0 \,.
\end{align}

The amplitude of the modulation is maximized in the face-on case, and each carrier's longitude $\phi$ contributes to the effective phase of the modulation. 
For the edge-on case, the amplitude depends on the longitude of the carrier, but the phase of the modulation is common among the galactic binaries.

\subsection{Total gravitational wave signal}

Considering the results so far, we can write the total gravitational wave signal produced by a network of Galactic binaries. 
Each binary produces a carrier wave contribution $h_i$ to the total measured strain $h$.
Here $i$ indexes over the $N$ Galactic binaries.
If we account for both the local and carrier terms of the modulation, in the limit that the modulating frequency $f_m$ remains constant during timescales $\sim d_i/c$ for all the Galactic binaries, we have
\begin{align}
    h_i(t) = & A_i \cos\left[2 \pi f_i t - \alpha_i - \frac{A_m F_i f_i}{f_m} \sin \gamma_i \sin(2 \pi f_m t -\delta_i) \right] 
    \notag \\
    \approx & 
    A_i \cos(2 \pi f_i t - \alpha_i) 
    \notag \\
    & + \frac{A_i A_m F_i f_i}{f_m} \sin \gamma_i \sin(2\pi f_i t - \alpha_i)\sin(2 \pi f_m t - \delta_i) \, ,
    \label{eq:hCarrierTime}
\end{align}
and the total signal arising from the sum of carriers is
\begin{align}
    h(t) & =\sum_i h_i(t) \,.
    \label{eq:hTotalTime}
\end{align}
In the opposite extreme, where $f_m$ evolves rapidly (but is still approximately constant over the timescale $T$ of the observations), we can neglect the effect of the carrier (pulsar) terms.
Then we can make the simple substitution $\gamma_i \to \pi/2$ and divide $A_m$ by an additional factor of two to get $h(t)$.
Specifically,
\begin{align}
    h_i(t)
    \approx & 
    A_i \cos(2 \pi f_i t - \alpha_i) 
    \notag \\
    & + \frac{A_i A_m F_i f_i}{2f_m} \sin(2\pi f_i t - \alpha_i)\sin(2 \pi f_m t - \alpha_m - \beta_i)
\end{align}
in this case.

\section{Gravitational Wave Timing Array Sensitivity}
\label{sec:Sensitivity}

We now turn to the problem of estimating the sensitivity of a network of nearly monochromatic carrier sources to modulation by a background, low-frequency gravitational wave.
We carry out the sensitivity estimate in three ways. 
First, we use time-domain techniques borrowed from the familiar method of time-delay measurements in pulsar timing arrays to estimate the sensitivity of the gravitational wave timing array.
We confirm this approach using frequency-domain methods, employing matched filtering to create simple signal-to-noise ratio (SNR) estimates of the sensitivity of the array.
Finally, we use a Fisher matrix approach to arrive at a more complete frequency-domain estimate for the sensitivity of a gravitational wave timing array, applicable at both high and low modulating frequencies.

\subsection{Timing sensitivity estimates}
\label{sec:TimeDomainSensitivity}

Although not as accurate as the frequency-based sensitivity approaches presented in subsequent sections, a timing estimate of the gravitational wave timing array sensitivity provides a good approximation to the more complete results, and illustrates the function of the array in a way most analogous to pulsar timing array methods.

We break up the full length of the observation period $T$ into a set of segments each with duration $\delta t$. 
Within each observation period of length $\delta t$, we consider the question of how well we can measure a time (or phase) delay in the arrival of gravitational waves from a Galactic binary. 
This time delay could be caused, for example, by a long-wavelength gravitational wave that is modulating the frequency of the carrier wave. 
Within a given time interval, this would appear as a small phase shift in the carrier signal, which would not be present if the carrier frequency were fixed. 
Let $\sigma_{t_\mathrm{d}}$ be the error with which we can measure such a time delay.

We model the gravitational wave signal from a single Galactic binary as a sinusoid with amplitude $A_i$ and frequency $f_i$, and we denote the induced time delay by $t_\mathrm{d}$.
This simplified model does not fully capture the signal from Galactic binaries in the presence of modulating waves and results in an optimistic sensitivity estimate. 
The signal waveform from the binary is given by
\begin{equation}
    h_i(t) = A_i \cos{[2 \pi f_i (t-t_\mathrm{d})]} \, .
\end{equation}
We assume that $\delta t$ is small enough that any drift in the frequency $f_i$ over the duration is negligible, and we treat the amplitude $A_i$ as constant during this window as well.

We approximate the noise on the observed gravitational wave strain as white noise in the neighborhood of the carrier frequency, since we treat $f_i$ as being nearly constant over the period $\delta t$. 
Using $S_n(f)$ to denote the one-sided power spectral density of the noise on the observed strain (e.g. from the LISA mission~\cite{LISAScienceReq}), we take $S_n(f_i)$ to denote the power spectral density of white noise with a fixed value of $S_n(f=f_i)$ for all $f$.

From here, we take a Fisher matrix approach, where the parameters $\theta^a$ that we would like to estimate from our observations are $t_\mathrm{d}$ and $f_i$. Then the elements of the Fisher matrix for a waveform given in the time domain are~\cite{Swerling1053694}
\begin{equation}
    \Gamma_{ab} = \frac{2}{S_n(f_i)} \int_{-\delta t/2}^{\delta t/2} \frac{\partial h_i(t, t_\mathrm{d}, f_i)}{\partial \theta^a} \frac{\partial h_i(t, t_\mathrm{d}, f_i)}{\partial \theta^b} \dd t \, .
    \label{eq:TimeDomainFisher}
\end{equation}
The lower bound on the variance of any unbiased estimator of these parameters can be found from the covariance matrix given by the inverse of the Fisher matrix $\Sigma_{ab} = \Gamma_{ab}^{-1}$. 
The Fisher matrix elements are straightforward to compute and are given by
\begin{align}
	\Sigma_{t_\mathrm{d} t_\mathrm{d}} & = \frac{1}{(2\pi)^2} \frac{S_n(f_i)}{A_i^2 f_i^2 \delta t} \,, \\
	\Sigma_{t_\mathrm{d} f_i} & = 0 \,, \\
	\Sigma_{f_i f_i} & = \frac{3}{\pi^2} \frac{S_n(f_i)}{A_i^2 \delta t^3} \,,
\end{align}
where we have assumed that the time delay $t_\mathrm{d}$ and the period of the Galactic binary carrier wave are small compared to $\delta t$. From here, we get the error in measuring the time delay,
\begin{equation}\label{eq:tsigma}
    \sigma_{t_\mathrm{d}} = \sqrt{\Sigma_{t_\mathrm{d} t_\mathrm{d}}} = \frac{1}{2\pi} \sqrt{\frac{S_n(f_i)}{A_i^2 f_i^2 \delta t}} \, ,
\end{equation}
given some observing ``cadence'' $1/\delta t$.

To determine the sensitivity of a gravitational wave timing array to the presence of a modulating wave, we must combine the information provided by timing measurements from $N$ binaries, each with a timing measurement error of $\sigma_{t_\mathrm{d}}$, while the time-based subdivision of the full data set provides a cadence $1/\delta t$. 
This is analogous to finding the sensitivity of a pulsar timing array whose measurements have a given timing error and cadence, and thus we can apply existing estimates for pulsar timing array sensitivity curves.  

The total strain sensitivity of a pulsar timing array may be approximated as the sum of two power laws, describing the low- and high-frequency limit, and given respectively by~\cite{Moore:2014eua}
\begin{equation}
    h_c^\mathrm{LOW}(f_m) \approx \frac{3\sqrt{\varrho_{\mathrm{th}}}}{2^{7/4}\chi\pi^3}\left(\frac{13}{N(N-1)}\right)^{1/4}\frac{\sigma_{t_\mathrm{d}}}{f_m^2 T^3} \sqrt{\frac{\delta t}{T}} \sec{\xi} \, ,
    \label{eq:MooreLowFreq}
\end{equation}
and
\begin{equation}
    h_c^\mathrm{HIGH}(f_m) \approx \left(\frac{16 \varrho_{\mathrm{th}}^2}{3\chi^4 N(N-1)}\right)^{1/4}\sigma_{t_\mathrm{d}} f_m\sqrt{\frac{\delta t}{T}} \, ,
        \label{eq:MooreHighFreq}
\end{equation}
where $\varrho_{\mathrm{th}}$ is the threshold SNR value above which a detection is claimed, $N$ is the number of pulsars in the array, $\chi$ is a geometric factor which is $1/\sqrt{3}$ for pulsar timing arrays, $T$ is the total time span of the observations, and $\xi$ is chosen such that $h_c^\mathrm{LOW} = h_c^\mathrm{HIGH}$ at a frequency of $2/T$~\cite{Moore:2014eua}. 
The spatially averaged geometric factor for the gravitational wave timing array is $\chi = \langle F \rangle$ given by $4/3$ for face-on binaries and $1/6$ for edge-on binaries. 
The characteristic strain $h_c$ corresponds, up to factors of order unity, with the amplitude of the modulating wave $A_m$ defined in Eqs.~\eqref{eq:hTT} and \eqref{eq:PolarizationAmps}.
We can see from Eq.~\eqref{eq:tsigma} that under the approximations made in deriving that result, the resulting gravitational wave timing array sensitivity curve does not depend explicitly on $\delta t$.

We apply this estimate to the gravitational wave timing array results with $N$ Galactic binaries included in the array, using the assumption of rapidly evolving sources so that we can neglect the pulsar term as in~\cite{Moore:2014eua}.
In order to apply this approximation to the gravitational wave timing array, we need to go beyond the simplifying assumption implicit in Eqs.~\eqref{eq:MooreLowFreq} and \eqref{eq:MooreHighFreq} to allow for different timing uncertainties $\sigma_{t_\mathrm{d}}$ for each binary.
In Sec.~\ref{sec:mock} we evaluate this sensitivity estimate using a mock catalog of LISA sources, where $\sigma_{t_\mathrm{d}}$ varies significantly among the binaries. 
We introduce a modification to Eqs.~\eqref{eq:MooreLowFreq} and \eqref{eq:MooreHighFreq}, making the replacement
 \begin{align}
 \label{eq:TimingSubstitution}
 \frac{\sigma_{t_\mathrm{d}}}{[N(N-1)]^{1/4}} 
 & \rightarrow \left[\sum_i \sigma_{t_\mathrm{d},i}^{-2} \right]^{-1/2} \,.
\end{align}
This replacement agrees in the limit of many identical binaries and weights the contribution of each binary according to its SNR in the expected manner.
However, it highlights another difference in the assumptions made in \cite{Moore:2014eua} and our work: Eqs.~\eqref{eq:MooreLowFreq} and \eqref{eq:MooreHighFreq} are derived using the cross-correlation among the pulsars in the network, whereas our uncertainty estimates arising from Eq.~\eqref{eq:TimeDomainFisher} come from an autocorrelation for each binary. 
As such, this replacement is heuristic, and should not be taken as a rigorous large-$N$ limit.

We emphasize that due to differences in conventions and definitions between our work and~\cite{Moore:2014eua}, we do not intend for this to be a precise mapping onto their results. 
Our timing estimate of the sensitivity is meant to illustrate the concept of the gravitational wave timing array, and particularly, in a way that is analogous to the measurements taken by pulsar timing arrays. 
It does not match precisely with our results from the more accurate frequency-domain approaches given in subsequent sections.
We compare this timing estimate to frequency domain estimates using a mock catalog of LISA detections in Sec.~\ref{sec:mock}.

\subsection{Matched filtering sensitivity estimate}
\label{subsec:MatchedFilter}

A common approach to the detection of gravitational waves is the use of matched filtering.
We assume that the noise in the gravitational wave detector is approximately colored Gaussian noise, characterized by a one-sided spectral noise density $S_n(f)$.
It is then natural to define a noise-weighted inner product between two frequency-domain signals $\tilde g(f)$ and $\tilde h(f)$,
\begin{align}
\label{eq:InnerProduct}
    \langle \tilde g | \tilde h \rangle 
    &= 4 \, {\rm Re} \int_{f_l}^{f_h} \frac{\tilde g^* \tilde h}{S_n(f)} \dd f \,.
\end{align}
Here $f_l$ is a lower frequency cutoff for the integral, and $f_h$ is an upper frequency cutoff which must be below the Nyquist frequency (half the sampling rate of the time series).
We assume we work with the Fourier transforms of real time-domain signals, so that the signals at negative frequencies are given by the complex conjugate of the signals at positive frequencies.
The prefactors of Eq.~\eqref{eq:InnerProduct} account for these facts, and in what follows we write only the positive-frequency parts of the signals $\tilde h$, with the negative parts implied.

The optimal linear detection statistic for a signal $\tilde h$ in the presence of such noise $\tilde n$ is the matched filter between data, $\tilde d = \tilde h + \tilde n$, and $\tilde h$. 
Its expectation value is the squared SNR (e.g.~\cite{Maggiore_2008}),
\begin{align}
    \rho^2 = \overline{\langle \tilde d | \tilde h \rangle } = \langle \tilde h | \tilde h \rangle \,,
\end{align}
where the overbar denotes expectation value.
Thus the product of the signal with itself gives the (squared) SNR that might be achieved by matched filtering.

We estimate the sensitivity of the gravitational wave timing array by isolating the sideband contributions to the gravitational waves of the array of monochromatic signals, and squaring this contribution.
We first treat the case where we include the carrier (pulsar) terms in the modulation.
Let $\tilde h(f)$ be the total gravitational wave signal; a finite-time Fourier transform over the period of observation applied to Eqs.~\eqref{eq:hCarrierTime} and \eqref{eq:hTotalTime} gives
\begin{align}
\tilde h(f) & = \sum_i \left[\tilde h_i(f) + \tilde s_i(f)\right]\,,
\end{align}
where the carrier wave contributions are
\begin{align} 
\tilde h_i & = \frac{A_i e^{i \alpha_i}}{2} \delta_T(f-f_i) \,, 
\end{align}
and around each carrier wave is a pair of sidebands whose contributions are
\begin{align}
 \tilde s_i  = \frac{A_i A_m F_if_i  \sin \gamma_i e^{i \alpha_i}}{4f_m} [ 
 & e^{-i \delta_i} \delta_T(f - f_i + f_m)
 \notag \\ & 
 - e^{i\delta_i} \delta_T(f - f_i - f_m) ] \,.
\end{align}
Here we write $\delta_T(f)$ for the finite-time delta function following~\cite{Moore:2014eua},
\begin{align}
    \delta_T(f)  = T \sinc(\pi f T)\,,
\end{align}
with $T$ the total observation time.
When $\pi f T \gg 1$, $\delta_T$ is highly peaked around $f = 0$. 
We emphasize again that we write only the positive-frequency parts of these signals. 

The squared SNR in the sidebands is
\begin{align}
    \rho_{\rm side}^2 & = 
    \sum_{ij} \langle \tilde s_i | \tilde s_j \rangle \,.
\end{align}
In evaluating this product, we consider observation times $T$ long compared to the periods of any of the carrier waves, so $\pi f_i T \gg 1$.
Then the products of sidebands around distinct carrier signals have negligible overlaps, so that the product vanishes when $i\neq j$.
Similarly, when $\pi f_m T \gg 1$ we expect each of the two sidebands around each carrier to be well separated, and the products of different sidebands vanish.
However, when we consider modulating waves with very long periods of order $T$, the two sidebands around each carrier can have nonzero overlap with one another and with the carrier peak.

It is convenient to first compute the squared SNR of a single carrier wave peak,
\begin{align}
    \rho_i^2 = \langle \tilde h_i | \tilde h_i \rangle 
    & = \frac{A_i^2}{4} \langle e^{i \alpha_i} \delta_T(f-f_i)|e^{i \alpha_i} \delta_T(f-f_i) \rangle
    \notag \\
    & \approx \frac{A_i^2 T}{\pi S_n(f_i)} \int_{-\infty}^{\infty} \sinc^2 x \, \dd x 
    = \frac{A_i^2 T}{S_n(f_i)} \,.
\end{align}
In going to the second line we made a coordinate transformation $x = \pi (f-f_i) T$, which centers the integral on the location of the carrier wave peak and stretches out a small region around that peak to a large domain.
This allows us to approximate $S_n(f)$ in the integral with its constant value at the peak, and extend the integral over an infinite range of $x$. 
We then used an integral identity for the square of a sinc function.

Using the same approximations, we compute
\begin{align}
   \langle & \delta_T(f-f_i+f_m) | \delta_T(f-f_i+f_m) \rangle 
    \notag \\
  &  \approx
    \frac{4 T}{\pi S_n(f_i)} \int_{-\infty}^{\infty} \sinc^2(x + x_m) \dd x = \frac{4 T}{S_n(f_i)}\,.
\end{align}
The overlap of two factors of $\delta_T(f-f_i - f_m)$ gives the same result, but the overlap of neighboring sidebands is
\begin{align}
   \langle & \delta_T(f-f_i+f_m) | \delta_T(f-f_i-f_m) \rangle 
    \notag \\
  &  \approx
    \frac{4 T}{\pi S_n(f_i)} \int_{-\infty}^{\infty} \sinc(x + x_m) \sinc(x-x_m) \dd x \,.
\end{align}
To resolve this we need the identity
\begin{align}
\label{eq:SincIdentity}
    \int_{-\infty}^{\infty} \sinc(x + a) \sinc(x + b) \dd x = \pi \sinc (a-b) \,.
\end{align}
With these results we find
\begin{align}
\label{eq:SidebandFinal}
     \rho_{\rm side}^2 
      \approx &
     \frac{A_m^2}{2 f_m^2}\sum_i (\rho_i F_i f_i \sin \gamma_i)^2 
     \notag \\ &
     - \frac{A_m^2}{2 f_m^2} \sinc(2\pi f_m T) \sum_i (\rho_i F_i f_i \sin \gamma_i)^2 \cos(2\delta_i) \,.
\end{align}
This shows that the squared SNR is the weighted square-sum of the individual carrier wave SNRs, so that the sideband SNR grows with $\sqrt{N}$.
The terms are enhanced by the factors $f_i^2/f_m^2$, which count the number of cycles of the carrier wave over which we accumulate the modulating signal.
At high modulating frequencies, $\pi f_m T \gg 1$, we can neglect the second term in Eq.~\eqref{eq:SidebandFinal}, and we see that $\rho_{\rm side}$ decreases as $1/f_m$, limiting  sensitivity at high modulating frequencies.
For a generic modulating wave, we also expect that $\delta_i$ varies randomly among the Galactic binaries, and so this second sum also tends to be suppressed by contributions with different signs.

We also expect the gravitational wave timing array loses sensitivity at very low modulating frequencies, a behavior that is not captured in Eq.~\eqref{eq:SidebandFinal}.
We return to this issue in Sec.~\ref{sec:Fisher} and use only the high-frequency approximation to Eq.~\eqref{eq:SidebandFinal} for the moment.

To get a sensitivity estimate, we can set a threshold SNR $\rho_{\rm th}$ above which we claim a detection of the sideband power. 
Then the required modulating amplitude for detection is
\begin{align}
	\label{eq:SNREstimate}
	& A_m  \gtrsim
	\rho_{\rm th} \frac{2}{\sqrt{N}} 
	\frac{f_m}{\rho_{\rm RMS} F_{\rm RMS}f_{\rm RMS}} \notag \\
	&	\sim 10^{-7}
		\bigg(\frac{\rho_{\rm th}}{1}\bigg)
		\bigg(\frac{f_m}{ \rm \mu Hz}\bigg)
		\bigg(\frac{100}{\rho_{\rm RMS}}\bigg)
		\bigg(\frac{1}{F_{\rm RMS}}\bigg)
		\bigg(\frac{\rm mHz}{f_{\rm RMS}}\bigg)
		\bigg(\frac{10^2}{ \sqrt{N} }\bigg)
		\notag
		\\ 
		& (f_m T \gg 1), 
\end{align}
where RMS is the root-mean-square value of the given quantity over the array, assuming these variables are uncorrelated with each other. 
Here we averaged over $\gamma_i$ assuming a uniform distribution of angles, appropriate for the case where we include the carrier (pulsar) contributions to the modulation.

Since we are in the high-frequency limit, it is more appropriate to assume that $f_m$ evolves rapidly enough that we can neglect the carrier term contributions to the sidebands.
In the case where we neglect the carrier terms, the sideband signal is then
\begin{align}
 \tilde s_i  = \frac{A_i A_m F_i f_i e^{i \alpha_i}}{8f_m} 
 [ & e^{-i (\alpha_m + \beta_i)} \delta_T(f - f_i + f_m)
 \notag \\ & 
 - e^{i(\alpha_m + \beta_i)} \delta_T(f - f_i - f_m) ] \,.
\end{align}
Making the appropriate substitutions and taking the high-frequency limit,
\begin{align}
\label{eq:SidebandFinal2}
     \rho_{\rm side}^2 
      \approx &
     \frac{A_m^2}{8 f_m^2}\sum_i (\rho_i F_i f_i)^2\,, & (f_m T \gg 1) \,.
\end{align}
Overall, the threshold modulating amplitude is increased by a factor $\sqrt{2}$ compared to the case including the carrier terms.
The network is slightly less sensitive, because the carrier terms do not (incoherently) contribute to the measured signal in this regime.

\subsection{Fisher matrix sensitivity estimate}
\label{sec:Fisher}

The sensitivity estimate using the sideband SNR threshold, Eq.~\eqref{eq:SNREstimate}, gives a reasonable result for modulating waves with high modulating frequencies.
However, our SNR estimate predicts ever-improving sensitivity at lower $f_m$, even for $f_m < 1/T$.
This is because the estimate assumes more information can be extracted from the signal than is actually available at these low frequencies.
In reality, uncertainties in the estimated parameters of the binaries in the array and in the modulating wave itself correlate with the uncertainty in the measured amplitude $A_m$, limiting our ability to distinguish $A_m$ from zero.

To capture these effects, we turn to a Fisher-based estimate of the measurement uncertainties of the gravitational wave timing array.
In this language, we can detect the modulating wave if the posteriors for $A_m$ peak sufficiently far from zero relative to their width.
To estimate the width, we need the relevant terms of the covariance matrix $\Sigma_{ab}$ for our signal model.
As in Sec.~\ref{sec:TimeDomainSensitivity} the covariance matrix is the inverse of the Fisher matrix $\Gamma_{ab}$, whose entries are given here by
\begin{align}
    \Gamma_{ab} & = \left \langle  \frac {\partial \tilde h}{\partial \theta^a} \right. \left | \frac{\partial \tilde h}{\partial \theta^b} \right \rangle\,,
\end{align}
where $\theta^a$ is the vector of model parameters for the array.

In addition, when considering the sensitivity of this array at low frequencies $f_m$, the effect of a slow drift in the carrier frequencies $f_i$ becomes important.
This is true when estimating the sensitivities of pulsar timing arrays, where the low-frequency sensitivity of Eq.~\eqref{eq:MooreLowFreq} is limited by the need to fit the pulsar spin down rate.
In binaries the frequency can change due to the slow gravitational-wave driven inspiral, and in the case of stellar binaries it can also occur due to tidal interactions and mass transfer~\cite{Willems:2007nq}.
To capture these effects, we allow for the slow evolution $\dot f_i$ of the carrier frequencies and compute the Fisher matrix including these parameters, but we take only the leading (zeroth) order results in small $\dot f_i$ for $\Gamma_{ab}$.
Our model for $h(t)$ and $\tilde h(f)$ when including nonzero $\dot f_i$ is described in Appendix~\ref{sec:DriftAppx}.

\subsubsection{Fast evolution of the modulating wave}

We first treat the case where the evolution of $f_m$ is fast enough that we can neglect the carrier terms, as discussed in Sec.~\ref{sec:FastEvolve}.
In this case, the model parameters are $\theta^a = \{ A_1, \alpha_1, f_1, \dot f_1, A_2, \dots, A_m, \alpha_m \}$.
The first $4N$ entries correspond to the parameters of the $N$ Galactic binaries, and the final two entries correspond to the parameters of the modulating wave.
For simplicity we do not incorporate the $F_i$ and $\beta_i$ terms into the Fisher analysis in this case, instead fixing the direction of propagation, polarization angle $\psi$, and inclination $\iota$ of the source of the modulation, and assuming that the sky locations of the binaries are relatively well measured.
For the modulating wave, we include only the modulating amplitude $A_m$ and common phase term $\alpha_m$ as parameters in the Fisher matrix, meaning that we marginalize over the phase to get a sensitivity estimate of the gravitational wave timing array at a fixed frequency $f_m$.

We compute the entries of the Fisher matrix using the same approximations as we did for $\rho_{\rm side}$, but the computations are more involved.
Some details and all the entries of $\Gamma_{ab}$ are given in Appendix~\ref{sec:FisherApp}.
Our goal is to compute the component of the covariance matrix 
$\Sigma_{A_m A_m}$, describing uncertainty in the measurement of $A_m$.
At high modulating frequencies, we find 
\begin{align}
    \Sigma_{A_m A_m} & \approx 8 f_m^2 \left[ \sum_i (\rho_i F_i f_i)^2 \right]^{-1} \,, & &  (f_m T) \gg 1 \,,
    \label{eq:hfreq_var}
\end{align}
in agreement with our SNR threshold estimate.
The dependence on the common phase $\alpha_m$ vanishes in this limit, which is expected\textemdash{}for times long compared to the period of modulation, the $\delta_T$ terms become very sharp, and one can check that a global time shift will remove the $\alpha_m$ phases from the gravitational-wave signal.
In the case where $f_m$ is smaller, we cannot execute this global shift, since the center of our observing window is used as our origin of time, and the relative phase of the modulating wave can change its impact on the carrier signals.

At low frequencies,
we find
\begin{align}
\label{eq:6dLowFreqSensitivity}
    \Sigma_{A_m A_m} \approx & \frac{396900}{(\pi T)^8 f_m^6} \left[\sum_i [\rho_i F_i f_i \csc(\alpha_m+\beta_i)]^2\right]^{-1} \,, 
   \notag \\ &  
   (f_m T) \ll 1 \,.
\end{align}
We see that there is a steep loss in sensitivity towards very small $f_m$, in agreement with expectations that the array is insensitive to very low frequency waves.
Note that the apparent singularities at $\delta_i = \alpha_m + \beta_i = 0$ and $\pi$ are artifacts of the expansion in small $f_m T$, which does not hold for these values of $\delta_i$. 
Instead, the power law becomes $f_m^{-4}$, shallower than $f_m^{-6}$, for these cases.

\begin{figure}[tb]
    \centering
    \includegraphics[width=\columnwidth]{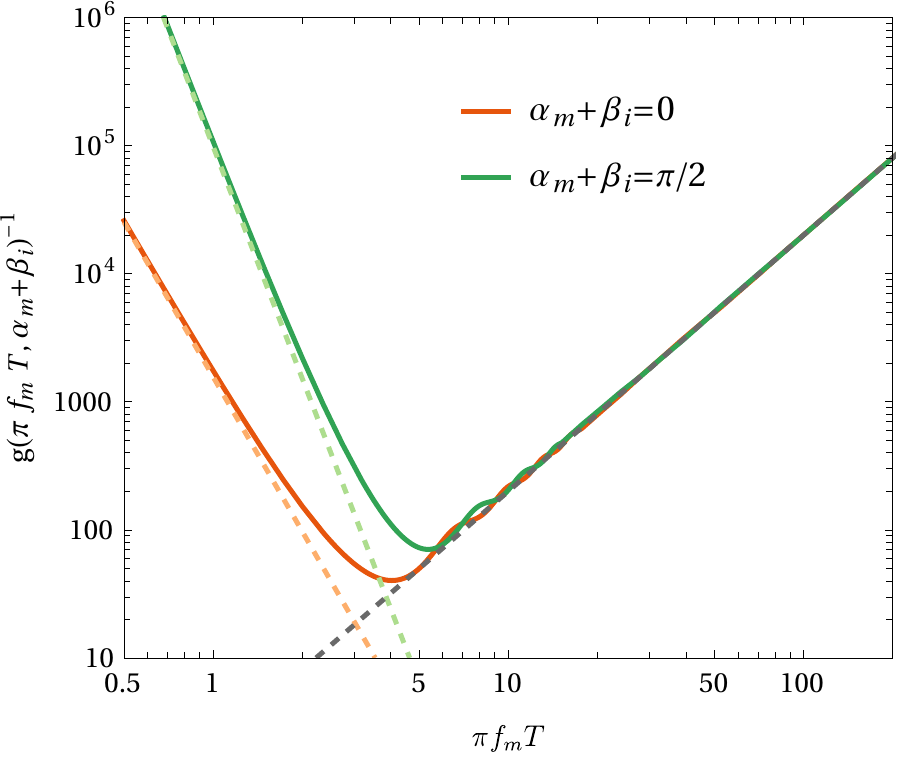}
    \caption{Generic shape of the variance of the amplitude of a modulating wave as a function of modulating frequency after marginalizing over \{$
    A_i,\alpha_i,f_i,\dot f_i,\alpha_m$\}, for the two limiting cases in which all carrier phases are chosen such that $\alpha_m+\beta_i=\pi/2$ or $\alpha_m+\beta_i=0$.
    The functional form of $g$ is given in Eq.~\eqref{eq:SensitivityShape}.
    The low- and high-frequency limits are shown with dashed lines. The low-frequency limit of the variance has the generic form $\propto f_m^{-6}$, with a different scaling of $f_m^{-4}$ when $\alpha_m+\beta_i$ approaches $0$ or $\pi$. The variance scales as $f_m^2$ in the high-frequency limit. 
    }
    \label{fig:analytic}
\end{figure}

The full expression for $\Sigma_{A_m A_m}$ is given by
\begin{align}
{\Sigma_{A_{m}A_{m}}} & = 
4 \left[\sum_i \rho_i^2 F_i^2 (\pi f_i T)^2 g(\pi f_m T,\alpha_m + \beta_i) \right]^{-1} \,,
\label{eq:fullsigmaAm}
\end{align}
where the definition of the function $g$ is given in Appendix~\ref{sec:FullSigmaAppx}, Eq.~\eqref{eq:SensitivityShape}, and is non-singular for all $\delta_i$.
In Fig.~\ref{fig:analytic} we plot $[g(\pi f_m T,\alpha_m+\beta_i)]^{-1}$, which controls the shape of the sensitivity curve of the array, as a function of the dimensionless parameter $\pi f_m T$.
We plot the extreme cases $\alpha_m+\beta_i = \pi/2$, which gives the generic power law behavior $\Sigma_{A_m A_m} \propto f_m^{-6}$, and $\alpha_m+\beta_i = 0$, which displays the shallower power law behavior at low $f_m$.
This sensitivity curve transitions from the high-frequency to low-frequency limit around the frequency where the sidebands plotted in Fig.~\ref{fig:freq_diag} begin to blend with the main peaks, which occurs when $f_m \sim 1/T$.

We can repeat our Fisher analysis while also marginalizing over $f_m$, which gives a conceptually different sensitivity estimate. Namely, rather than an estimate of the sensitivity to a fixed frequency of gravitational waves, it incorporates the uncertainties inherent in a search over modulating frequencies. 
In this case, the high-frequency sensitivity estimate of Eq.~\eqref{eq:6dLowFreqSensitivity} remains unchanged, while the low-frequency behavior becomes steeper, with $\Sigma_{A_m A_m} \propto f_m^{-8}$.

\subsubsection{Slow evolution of the modulating wave}
\label{sssec:slowevol}

When the modulating frequency evolves sufficiently slowly, the sidebands produced by the local term and carrier terms overlap.
In this case, the carrier terms also contribute to the modulating signal, and two complications arise in the Fisher analysis.
First, the phases $\delta_i$ are no longer determined by the common phase $\alpha_m$ and a function of sky location $\beta_i$. 
Instead, they depend on the unknown distance to each Galactic binary $d_i$ through the phase $\gamma_i$.
Secondly, the amplitude of the modulation also varies with $d_i$, due to the presence of the factors $\sin \gamma_i$; see Eq.~\eqref{eq:phase}.

In this case we must consider an expanded vector of parameters given by
$\theta^a = \{ A_1,\alpha_1,f_1,\dot{f}_1,\delta_1, A_{m1}, A_2, \dots \}$, where we have defined $A_{mi} = A_m \sin \gamma_i$.
So long as we treat the common modulating frequency $f_m$ as a fixed parameter, the resulting Fisher matrix is block diagonal, with a $6 \times 6$ block for each carrier.
We can therefore invert each block independently to find the uncertainties $\sigma_{A_{mi}}$ on each of the $A_{mi}$.
The resulting expression for $\sigma_{A_{mi}}^2$ is similar to the expression for $\Sigma_{A_m A_m}$ in Eq.~\eqref{eq:fullsigmaAm}, but without the sum over binaries.
This means at high frequencies,
\begin{align}
    \sigma_{A_{mi}}^2 & \approx 2 f_m^2 (\rho_i F_i f_i)^{-2} \,,
    & (f_m T) \gg 1 \,,
\end{align}
and at low frequencies
\begin{align}
    \sigma_{A_{mi}}^2 & \approx \frac{99225}{(\pi T)^8 f_m^6}[\rho_i F_i f_i \csc(\delta_i)]^{-2} \,,
    & (f_m T) \ll 1\,. 
\end{align}
The full expression, valid for all frequencies and values of the phases $\delta_i$, is given in Appendix~\ref{sec:FullSigmaAppx}, Eq.~\eqref{eq:SlowSigmaAm}.

In order to compute the sensitivity of the whole array to a common modulating wave with amplitude $A_m$, we must combine the sensitivities from each of the carriers.
We have observations of $N$ individual $A_{mi}$, and we would like to determine the error on $A_m$, given that $A_{mi}=A_m \sin \gamma_i$.
However, this constitutes $N$ measurements with which to constrain $N+1$ correlated parameters, and it is straightforward to see that the Fisher matrix including $A_m$ and all of the $\gamma_i$ is singular.
This suggests that there is no unbiased estimator with finite variance for the quantity $A_m$ (without imposing additional constraints).
This is due simply to the fact that a very large value of $A_m$ could always be compensated by some choice of the $\gamma_i$ to produce the same data.
On the other hand, it is still possible to estimate our sensitivity to detecting the presence of a modulating wave.
One procedure to do so is to compute the Moore-Penrose pseudo-inverse of the singular Fisher matrix, which would allow us to find the constrained Cramer-Rao bound on the variance of the modulating amplitude~\cite{Li_2012}.

We will proceed by a different route, and utilize the likelihood ratio test to compare a model that accounts for the presence of a modulating wave with amplitude $A_m$ and a set of phases $\gamma_i$ to the null hypothesis with no parameters.
Under the null hypothesis, any apparent modulation results purely from noise.
The log-likelihood ratio for these two models is given by
\begin{align}
    D = -2 \ln\left(
    \frac{\mathcal{L}_0}
    {\max_{\theta}\mathcal{L}(\theta)}
    \right) \, ,
    \label{eq:LogLikeRatio}
\end{align}
where we estimate the likelihood as a Gaussian
\begin{align}
    \mathcal{L} = \prod_i 
    \frac{1}{\sqrt{2\pi} \sigma_{A_{mi}}}
    \exp\left(-\frac{(A_{mi} + n_i - A_m \sin\gamma_i)^2}{2\sigma_{A_{mi}}^2}\right) \, ,
    \label{eq:GaussLike}
\end{align}
where $n_i$ is the Gaussian noise with variance $\sigma_{A_{mi}}^2$ in the measurement of each $A_{mi}$.
The null hypothesis has $A_m = 0$.  
Plugging this in to the likelihood ratio, and maximizing the likelihood in the modulation model, we find
\begin{align}
    D = \sum_i \frac{\left(A_{mi} + n_i \right)^2}{\sigma_{A_{mi}}^2} \, ,
    \label{eq:LogLikeRatio2}
\end{align}
which, according to Wilks' theorem, is asymptotically $\chi^2$ distributed with $N+1$ degrees of freedom~\cite{Wilks_1938}.  The expectation value of this likelihood ratio when a modulating wave is present is given by
\begin{align}
    \langle D \rangle &= \left\langle \sum_i \frac{\left(A_m\sin\gamma_i + n_i\right)^2}{\sigma_{A_{mi}}^2} \right\rangle 
    = N + \sum_i\frac{A_m^2}{2\sigma_{A_{mi}}^2} \, ,
\end{align}
where we have used the fact that the $\gamma_i$ are uniformly distributed and taken the average over the network, $\langle\sin^2\gamma_i\rangle=1/2$.

Now since $D(A_m)\sim\chi^2(N+1)$, after choosing a detection threshold we can write the sensitivity of the array as the value of $A_m$ that exceeds the threshold.  
In the limit of large $N$, the quantity $\sqrt{2D(A_{m})}$ is approximately Gaussian distributed,
$\sqrt{2D(A_{m})} \sim \mathcal{N}(\sqrt{2N+1},1)$.  We can therefore approximate the threshold value of the modulating amplitude $A_{m,\mathrm{th}}$ for a $j$-$\sigma$ detection as
\begin{align}
    A_{m,\mathrm{th}} \approx \left[\frac{j^2 + 2j\sqrt{2N+1} + 1}{\sum_i \sigma_{A_{mi}}^{-2}}\right]^{1/2} \, .
    \label{eq:AmTh_GaussianApprox}
\end{align}
We find Eq.~\eqref{eq:AmTh_GaussianApprox} to be a very good numerical fit to the threshold derived when using the $\chi^2(N+1)$ distribution for $D(A_m)$, even for small $N$. 
As such, we use this more transparent expression in Sec.~\ref{sec:mock} when deriving the sensitivity of the array in the limit of a slowly evolving modulation.

Note that for an array of equally constraining binaries, where all $\sigma_{A_{mi}}$ are equal, the threshold $A_{m,\mathrm{th}}$ scales roughly as $N^{-1/4}$, as is apparent from Eq.~\eqref{eq:AmTh_GaussianApprox} in the large-$N$ limit.
However, when the $\sigma_{A_{mi}}$ are taken to be different for each binary, the addition of poorly measured binaries to the array can lead to an increased detection threshold by increasing $N$ (and thus the number of degrees of freedom of the modulation model) without a significant change to $\sum_i \sigma_{A_{mi}}^{-2}$.
In practice, the maximum sensitivity of the array in the case of slow evolution of the modulating wave is obtained by retaining only the binaries that provide the most significant contribution, those with the largest values of $ \rho_i f_i$.

\section{Gravitational Wave Timing Array Using Mock LISA Catalog}
\label{sec:mock}

With the analytic sensitivity estimates in hand, we can calculate the sensitivity of a gravitational wave timing array based on LISA measurements of gravitational waves from Galactic white dwarf binaries.
The binary white dwarf population in the Milky Way has been estimated to number on the order of millions. 
It is expected that data from LISA will be sufficient to individually resolve the gravitational waves from about $10^4$ white dwarf binaries with SNR~$>7$ over its planned four-year mission. 
Most of these binaries would be very nearly monochromatic for the extent of the LISA mission. 
The main source for their frequency evolution is likely to be gravitational wave radiation, with tides and mass transfer being effects only dominant in the later stages of inspiral (at frequencies above 1~mHz~\cite{Willems:2007nq}).

In order to have realistic inputs for white dwarf binary parameters and their SNR as measured by LISA over a four-year mission, we used the \emph{Radler} dataset~\cite{radler} created for the LISA Mock data challenges, in combination with the \texttt{Gbfisher} code~\cite{ldasoft}. \texttt{Gbfisher} computes the spacecraft ephemerides during the observation time and, in combination with input sources, creates a time delay interferometry data stream for the mission. 
After estimating the confusion noise, it proceeds to give a matched-filter SNR for each source.
In this analysis, we only use white dwarf binaries that \texttt{Gbfisher} identifies with SNR~$>7$ over a four-year LISA mission.
The median error on the sky localization of these binaries is  3 square degrees, which justifies our assumption that localization errors are negligible in our analysis.

For each sensitivity estimate, we calculate the modulating amplitude needed for a $3$-$\sigma$ detection. 
We denote this threshold as $A_{m,\rm{th}}$.
The timing estimate of the gravitational wave timing array sensitivity curve is calculated as the sum of Eqs.~\eqref{eq:MooreLowFreq} and \eqref{eq:MooreHighFreq}, using the timing residuals given by Eq.~\eqref{eq:tsigma}, with $\xi$ chosen to match the contributions at $f_m=2/T$.
In order to correctly weight the binaries contained in the mock LISA catalog, we implement the substitution shown in Eq.~\eqref{eq:TimingSubstitution} as described in Sec.~\ref{sec:TimeDomainSensitivity}. 
We use a detection threshold of $\varrho_{\rm th} = 3$ to roughly match the $3$-$\sigma$ criterion, and we average over the two limits of face-on and edge-on inclinations.

For the frequency-domain analysis, as stated in Sec.~\ref{sec:Fisher}, we consider two extreme cases for the evolution of the modulating wave: the fast case, when we can neglect the carrier term, and the slow case, when the carrier term has the same frequency as the local term.
In the case of fast evolution, we use the Fisher estimate for $\sigma_{A_m}$, given in full by Eq.~\eqref{eq:fullsigmaAm}, and set $A_{m,{\rm th}} = 3 \sigma_{A_m}$. 
We average over the phases $\delta_i$, as well as the properties of the modulating wave: its inclination $\iota$, polarization angle $\psi$, and direction of propagation $\hat k$.
For the case of slow evolution, we use the likelihood ratio threshold of Eq.~\eqref{eq:AmTh_GaussianApprox}, with $j=3$ and the full expression for $\sigma_{A_{mi}}$ from Eq.~\eqref{eq:SlowSigmaAm}.
We average over the same parameters as for the fast case.
Furthermore, as discussed in Sec.~\ref{sssec:slowevol}, in the slow case we select only the binaries with the highest $\rho_i f_i$, since including poorly measured binaries would worsen the forecasted sensitivity of the array.
We find that using the ten best binaries from the mock catalog gives the optimal estimate, and discuss this choice further below.

\begin{figure}[tb]
    \centering
    \includegraphics[width=\columnwidth]{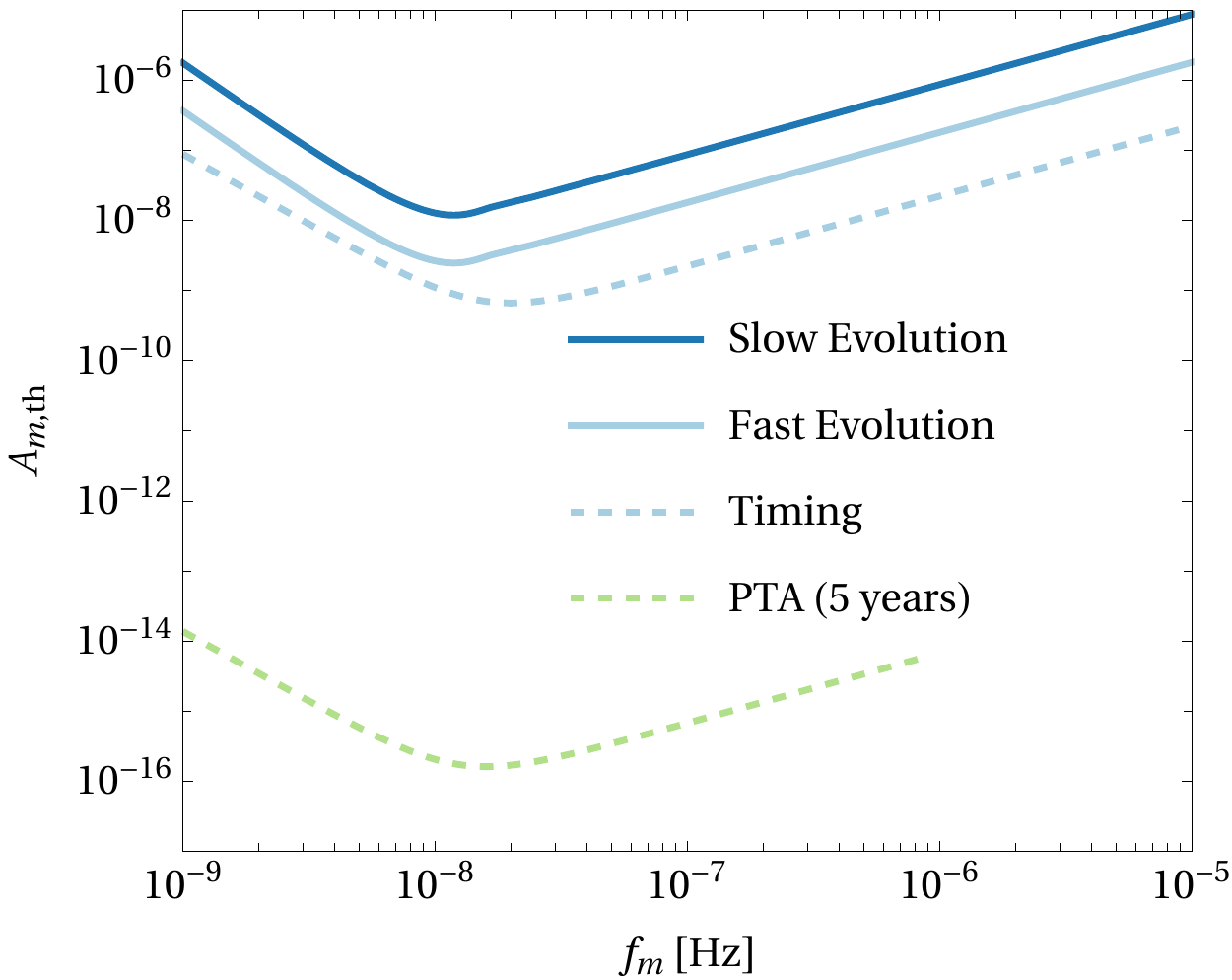}
    \caption{
    Forecasted sensitivity curves for a gravitational wave timing array based on the mock LISA catalog assuming a four-year mission.
    For the frequency-domain estimates, we show the case where the modulation has a slow evolution in dark blue and the case of fast evolution in light blue.
    The approximate sensitivity in the fast-evolution case from the timing estimate is shown in dashed light blue.  For comparison, we also plot the sensitivity of a pulsar timing array (PTA) assuming 5 years of monitoring $36$ pulsars with timing error $100$~ns in dashed light green.}
    \label{fig:results}
\end{figure}

Figure~\ref{fig:results} shows the estimated sensitivity curve for a gravitational wave timing array based on LISA observations given by each treatment.
We also include for comparison an estimate of pulsar timing array sensitivity, which is calculated from Eqs.~\eqref{eq:MooreLowFreq} and \eqref{eq:MooreHighFreq} assuming an array of 36 pulsars timed once per fortnight, with a total observation baseline of 5 years, a timing precision of 100~ns, and a sky-averaged geometric factor $\chi=1/\sqrt{3}$; this offers a direct comparison with results from~\cite{Moore:2014eua}.
In dashed light blue is the timing sensitivity estimate of the gravitational wave timing array.
The light and dark blue correspond respectively to the fast- and slow-evolution case evaluated with the mock catalog.

The timing estimate of the gravitational wave timing array shows a higher sensitivity than the more complete frequency-domain estimates. 
This higher sensitivity is to be expected, since the timing estimate is an order of magnitude calculation and includes fewer parameters that correlate with the amplitude of the modulation.
In the regime where both the gravitational wave timing array and pulsar timing arrays have sensitivity, the pulsar timing array is about 7 orders of magnitude more sensitive.
However, the sensitivity of the gravitational wave timing array extends above the maximum frequency of the pulsar timing array, all the way to the lower end of the LISA band ($\sim2\times10^5$ Hz).

\begin{figure}[tb]
    \centering
    \includegraphics[width=\columnwidth]{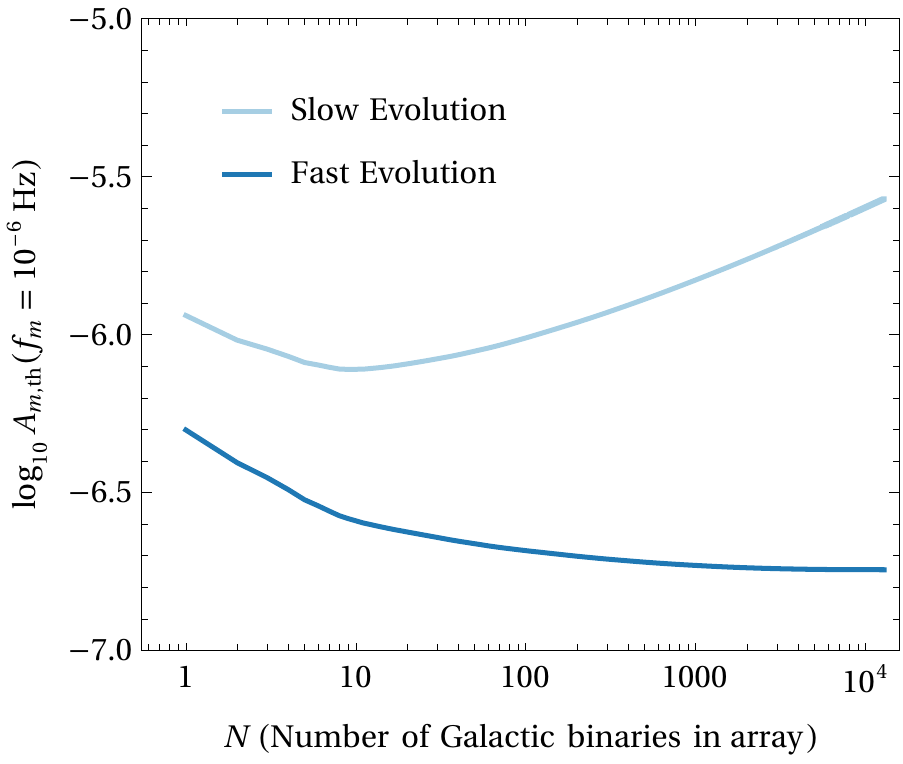}
    \caption{
    Dependence of the sensitivity of a gravitational wave timing array on the number of Galactic binaries used in the array, using a mock catalog of LISA observations~\cite{radler,ldasoft} assuming a four-year mission.
    The binaries are ordered by decreasing $\rho_i f_i$, which controls the contribution of each binary to the sensitivity. We use a \mbox{$3$-$\sigma$} detection threshold at a characteristic frequency of $10^{-6}$~Hz, and average over \{$\iota,\psi,\hat k$\}.
    In the slow-evolution case, adding ``noisy'' observations degrades the sensitivity, and we find that utilizing only the best 10 binaries gives the most sensitivity.
    In the fast-evolution case, the best 10 binaries (about $0.08\%$ of the total resolved population) can account for $\sim 40\%$ of the constraining power on $A_m$, but adding binaries always improves the sensitivity for the fast-evolution case.}
    \label{fig:N_Scaling}
\end{figure}

The sensitivity of the gravitational wave timing array can be improved in various ways, including via longer observation times or by monitoring a larger number of Galactic binaries with lower detector noise.
With a fixed number of binaries in the array, the threshold $A_{m,{\rm th}}$ scales with observation time as $T^{-1/2}$ due to the linear increase in $\rho_i^2$ with mission time.
In general this provides a lower bound on the improvement, since additional Galactic binaries will be resolved above the SNR threshold with increased observational time and can thus be added to the array.

Figure~\ref{fig:N_Scaling} shows how $A_{m,{\rm th}}$ changes with the number of detected binaries in each of the frequency-domain estimates, starting from those with the highest $\rho_i f_i $.
We use the standard four-year mission duration and plot $A_{m,{\rm th}}$ at a frequency $f_m=10^{-6}$ Hz (the high-frequency limit).
For the fast-evolution case, we use the high-frequency approximation for the measurement error, so that $A_{m,{\rm th}} \propto \left(\sum_i \rho_i^2 F_i^2 f_i^2 \right)^{-1/2}$, and again average over $\{\iota,\psi,\hat k$\}.
As can be seen in the figure, as few as 10 binaries (about 1/1000 of the total resolved population) can account for almost half of the sensitivity of the gravitational wave timing array in this case.
The threshold continues to improve with $N$ even when binaries with low $\rho_i f_i$ are included, but those with the highest $\rho_i f_i $ contribute significantly more to the sensitivity of the array.
In the slow-evolution case, the sensitivity decreases when more than the 10 best binaries are included, and so we use only these best 10 in the corresponding sensitivity curve of Fig.~\ref{fig:results}.

\section{Discussion}\label{sec:discussion}

In this paper we have discussed the concept of a gravitational wave timing array.
The array is composed of a large number of Galactic binaries, observed continuously through their gravitational wave emission by future detectors.
Lower frequency gravitational waves impact the propagation of the signals from these binaries, modulating their phase.
This effect can be used to search for these low-frequency gravitational waves in a matter directly analogous to pulsar timing array searches for gravitational waves.

We have specialized our results to the gravitational waves emitted by binary white dwarfs as measured by LISA, using a mock galaxy catalog and a mock LISA pipeline to estimate their properties.
In this case, the array provides sensitivity in the microhertz frequency regime, 
as illustrated in Fig.~\ref{fig:frange}.
Our main analysis is in the frequency domain, although we have also included an approximate timing estimate to allow for direct comparison with the standard pulsar timing array approaches.
Our results show that a gravitational wave timing array based on the nominal four-year LISA mission lifetime would provide a sensitivity to long-wavelength gravitational waves that is about 7 orders of magnitude worse than pulsar timing arrays in the overlapping region. 
Despite being a much lower sensitivity, an entirely independent probe into the nanohertz gravitational wave sky may prove useful, since it is unaffected by some of the major sources of noise for pulsar timing arrays, including electromagnetic propagation effects in the interstellar medium and radio interference at Earth.
The gravitational wave timing array also requires no change of design for the LISA mission, since it can be achieved by analyzing data that is already planned to be collected.

It is worth considering whether LISA or pulsar timing arrays could measure the microhertz regime directly, since their strain sensitivity is many orders of magnitude better than would be achieved by the indirect analysis with a gravitational wave timing array. 
As currently designed, the low-frequency sensitivity of LISA is significantly limited by acceleration performance, which effectively forms a frequency cutoff for the mission at \mbox{$\sim 10^{-4}$\textendash{}$10^{-5}$}~Hz~\cite{Larson2005,Babak:2021mhe}.
Pulsar timing arrays can extend their sensitivity into the microhertz regime by timing pulsars at a higher cadence than $\sim2$ weeks, as demonstrated by~\cite{Perera:2018pts}, achieving a strain sensitivity of better than $\sim 10^{-12}$ at $10^{-6}$ Hz.
Staggered observation of a large number of pulsars would also provide sensitivity to gravitational waves with frequency of several microhertz~\cite{Wang:2020hfh}.
An interesting recent proposal demonstrated that high-cadence photometric observations may be useful for astrometric gravitational wave searches in the microhertz regime~\cite{Wang:2020pmf}.
Relative astrometry of stars in the Galactic bulge with the Nancy Grace Roman Space Telescope~\cite{2019JATIS...5d4005W} may be capable of providing sensitivity to gravitational waves in the microhertz regime.
High precision monitoring of orbital dynamics is another avenue to search for microhertz gravitational waves~\cite{Blas:2021mpc,Blas:2021mqw}.
In any case, the gravitational wave timing array provides a valuable independent probe in the microhertz regime, using data products which the LISA mission already plans to produce. 

Our sensitivity estimates could be improved by including a more complete analysis of the effect of the carrier term (the pulsar term in pulsar timing array literature) on our sensitivity, possibly by modeling the frequency evolution of the modulating wave explicitly.
We do incorporate the presence of a slow evolution of the frequency of each binary in the array, whether it comes from gravitational-wave driven inspiral or from effects such as mass transfer or tides.
Improvements could also be made with a more realistic observing scenario for the LISA mission, taking into account non-stationary noise~\cite{2020PhRvD.102h4062E} and the mission's duty cycle, as well as the need to demodulate the orbital motion of the satellites from the raw data.

While we focused on the sensitivity of the array to a coherent source of low-frequency gravitational waves, it would be natural to also provide sensitivity estimates for a stochastic background of low-frequency gravitational waves, bursts of gravitational waves, and gravitational wave memory.
It would also be interesting to consider the impact of additional sources of sidebands in the binaries observed by LISA, such as orbital eccentricity, or interactions with a companion star in triple systems.
These effects would not generate sidebands at a common spacing for all the members of the gravitational wave timing array, but they may contribute an important source of noise for some of the binaries.

Although we have concentrated on the idea of a gravitational wave timing array using LISA observations, the idea may be applied to other situations.
For example, the detection of continuous wave sources by ground-based detectors (see e.g.~\cite{Riles:2017evm}), such as pulsars with some small equatorial ellipticity, would open up an additional avenue.
Searches for common modulations in a collection of such sources could benefit from the higher frequencies $f_i$ of the carrier waves, since the sensitivity scales as $f_i/f_m$, and could potentially access the entire range of sub-Hz frequencies not directly accessible to ground-based detectors.
Previous work has shown that observations of gravitational waves from cosmological neutron star binaries with a future observatory like BBO or DECIGO could provide sensitivity to a background of very low frequency gravitational waves with $f_m<10^{-12}$~Hz~\cite{Seto:2005tq}.

Gravitational wave astronomy is a rapidly evolving field that has already provided important insights that would not have been possible with electromagnetic observations alone.
New direct detection experiments and indirect detection schemes are under development that will greatly expand the frequency coverage and sensitivity with which we can search for gravitational waves.
The gravitational wave timing array described here is a novel proposal that offers a nearly cost-free extension of the sensitivity of future detectors to a currently unconstrained regime of gravitational wave frequencies.

\section*{Acknowledgments}
MJBR thanks Tyson Littenberg for his introduction to \texttt{Gbfisher} and the \textit{Radler} dataset.
MJBR and AZ are supported by NSF Grant PHY-1912578.
JM and CT are supported by the US Department of Energy under grant no.~DE-SC0010129.
NP was supported by the Hamilton Undergraduate Research Scholars Program at SMU.

\appendix 

\section{Modulation of a continuous signal}
\label{sec:GeometricOptics}

In this appendix we briefly discuss the geometric optics formalism required to derive the phase modulation of a monochromatic gravitational wave propagating in the presence of another, lower frequency gravitational wave.
The high-frequency carrier wave propagates in a perturbed flat spacetime $g_{\mu\nu} = \eta_{\mu \nu} + h_{\mu\nu}$ and is itself a tensor perturbation $\gamma_{\mu\nu}$.
We use a gauge such that $h_{\mu\nu}$ is transverse and traceless with propagation vector $k^\mu$.
The carrier wave is emitted at a stationary source, and received by a stationary observer at the origin of our coordinate system. 

The curvature scale of the perturbation is given by the wavelength of the modulating wave $\mathcal L \sim \lambda_m$, and so $\lambda_m \gg \lambda_e$.
This is the geometric optics limit, where high-frequency waves propagate along null geodesics in the curved spacetime, regardless of whether they are scalar, vector, or tensor fields.
The amplitude of these waves decreases as the wavefront expands, in order to conserve quanta (see e.g.~\cite{Misner:1973prb}).

More precisely, we augment our gauge such that $\gamma_{\mu \nu}$ is in Lorenz gauge and traceless with respect to $g_{\mu \nu}$, so that $g^{\alpha \beta} \gamma_{\alpha \beta}=0$ and $\nabla_\alpha \gamma_{\mu}{}^{\alpha} = 0$.
Then in vacuum $\gamma_{\mu \nu}$ obeys the wave equation
\begin{align}
\nabla^\alpha \nabla_\alpha \gamma_{\mu\nu} = 0 \,.
\end{align}
To implement the geometric optics limit we expand the carrier wave as
\begin{align}
\gamma_{\mu \nu} = \mathcal A e_{\mu \nu} e^{-i \varphi/\epsilon} \,,
\end{align}
where the phase function $\varphi$ varies rapidly over scales $\mathcal L$, the polarization tensor is normalized as $e_{\alpha \beta} e^{\alpha \beta} = 2$, and $\epsilon$ is a bookkeeping parameter which tracks orders in the rapidly varying phase.
The wave equation and gauge conditions are then solved order by order in the parameter $\epsilon$.
At leading order, the result is
\begin{align}
g^{\alpha \beta} (\partial_\alpha \varphi )(\partial_\beta \varphi) & = 0 \,, & 
g^{\alpha \beta} (\partial_\alpha \varphi) \nabla_\beta (\partial_\mu \varphi) = 0\,,
\end{align}
where the second equation is arrived at by differentiating the first and commuting the derivatives on the scalar $\varphi$.
If we define the normal vector to the wavefronts of constant phase as $\zeta_\mu = \partial_\mu \varphi$, we see that $\zeta^\mu$ is a null geodesic in the perturbed flat space, along which the carrier wave propagates.

From here, we can quote the standard results for such a null geodesic, perturbed by the modulating wave $h_{\mu \nu}$, see e.g.~Refs.~\cite{Book:2010pf,Anholm_2009}.
We expand $\varphi = \varphi_0 + \varphi_1 + \dots$ and so $\zeta^\mu = \zeta_0^\mu + \zeta_1^\mu + \dots$, counting orders in the small modulating wave amplitude.
If the spatial normal from the observer to the carrier source is $\hat n^i$, we have
\begin{align}
\zeta_0^\mu & = \omega_c (1, - \hat n^i) \,,
\end{align}
and $\zeta^\mu_1$ given by Eq.~(16) of~\cite{Book:2010pf}. 

The observed frequency at the origin is given by contracting the observer's 4-velocity $u^\mu = \delta_t^\mu$ with the phase derivative 
\begin{align}
\left. \frac{d\varphi}{dt}\right|_{\rm obs} \approx - u^\alpha g_{\alpha \beta} (\zeta^\beta_0 + \zeta^\beta_1) 
=\omega_c(1 - z) \,,
\end{align}
where the redshift $z$ is quoted in Eq.~\eqref{eq:redshift}.
This expression is used to compute the modulated phase $\varphi$.

\section{Including the slow evolution of the carrier frequencies}
\label{sec:DriftAppx}

In this appendix we consider the possibility that the gravitational waves from Galactic binaries are not purely monochromatic, and allow for a small linear time dependence in their instantaneous frequency, 
so that the carrier phase expands as
$\varphi_i = -\alpha_i + 2 \pi f_i t + \pi \dot f_i t^2$.
This consideration is motivated both by the fact that we expect some non-negligible frequency drift over the lifetime of our observations and also that we expect low frequency modulating waves to be degenerate with the frequency drift (as in the case of pulsar timing arrays).

When we include the slow evolution of the carrier frequencies, the modulated signal from a single Galactic binary is
\begin{align}\label{eq:ht_fdot}
    h(t)  = & A_i \cos\bigg[2 \pi f_i t + \pi \dot{f_i} t^2 -\alpha_i  \nonumber \\
    &\qquad \quad - \frac{A_m F_i f_i}{f_m} \sin \gamma_i \sin(2 \pi f_m t -\delta_i) \bigg] \nonumber \\
    \approx & 
    A_i \cos(2 \pi f_i t - \alpha_i) -  \pi \dot{f}_i t^2 A_i \sin(2 \pi f_i t - \alpha_i) \notag \\
    & + \frac{A_i A_m F_i f_i}{f_m} \sin \gamma_i \sin(2 \pi f_m t - \delta_i) \notag \\
    & \quad \times \bigg[ \sin(2\pi f_i t - \alpha_i)  + \pi \dot{f}_i t^2  \cos(2\pi f_i t - \alpha_i) \bigg] \,.
\end{align}
In the frequency domain, the carrier wave and sideband signals are then
\begin{align}
    \tilde h_i & = \frac{A_i e^{i \alpha_i}}{2} \left[ \delta_T(f-f_i) 
    + i \pi \dot{f_i} \delta_T''(f - f_i) \right] \,,
    \label{eq:hf_fdot}  \\
    \tilde s_i & =  \frac{A_i A_m F_if_i  \sin \gamma_i e^{i \alpha_i}}{4f_m} \notag \\
    & \times \left[ 
    e^{-i \delta_i} \left( \delta_T(f - f_i + f_m) + i \pi \dot{f}_i \delta_T''(f - f_i + f_m) \right) \right. \notag \\
    & \quad \left. - e^{i \delta_i} \left( \delta_T(f - f_i - f_m) + i \pi \dot{f}_i \delta_T''(f - f_i - f_m) \right)  \right] \,. 
    \label{eq:sf_fdot}
\end{align}
Since we work to leading order in the slow frequency evolution, we generally set $\dot f \approx 0$ in our final expressions.
This means we only need to consider the additional terms involving $\dot f_i$ when taking derivatives with respect to those parameters,
\begin{align}
    \partial_{\dot{f}_i} \tilde{h}_i &= i \pi \frac{A_ie^{i\alpha_i}}{2}   \delta_T''(f - f_i)  \, .
    \label{eq:hf_fdot_deriv}
\end{align}
These results are needed for approximating entries in the Fisher matrix $\Gamma_{a \dot{f}_i}$, as described below.

\section{Details on the Fisher matrix approach}
\label{sec:FisherApp}

In this appendix we give more detail on the Fisher matrix approach for estimating the sensitivity of the gravitational wave timing array.

\subsection{Slow evolution of modulating wave}

We treat first the case where both the local term (the Earth term in the pulsar timing array literature) and carrier  term (the pulsar term in the pulsar timing array literature) contribute to the sidebands with the same frequency offsets (but differing phases).
We define $A_{mi} = A_m \sin \gamma_i$, and our parameter set is $\theta^a = \{A_1,f_1,\alpha_1,\dot f_1, \delta_1, A_{m1}, A_2, \dots \}$.
We hold $f_m$, the sky locations of the binaries, and the other properties of the incident modulating wave fixed.
The resulting Fisher matrix is block diagonal, with one $6\times6$ block for each Galactic binary.
To leading order in the small $A_{mi}$, the entries in each block are given by
\begin{align}
	\Gamma_{A_i A_i} & = \frac{\rho_i^2}{A_i^2} \,, \\
	\Gamma_{\alpha_i \alpha_i} & = \rho_i^2\,, \\
	 \Gamma_{f_i f_i} & 
    =\frac{(\pi T)^2 \rho_i^2 }{3} \,, \\
    \Gamma_{\dot{f}_i \dot{f}_i}  &
     = \frac{\pi^2}{5} (\pi T )^4  \rho_i^2  \, ,\\
    \Gamma_{\delta_i \delta_i} & 
= \frac{A_{mi}^2 \rho_i^2  F_i^2 f_i^2}{2 f_m^2}[1 + \cos(2\delta_i)  \sinc(2 \pi f_m T)] \,, \\
    \Gamma_{A_{mi} A_{mi}} & = \frac{\rho_i^2 F_i^2 f_i^2}
    {2 f_m^2} 
    [1 -\cos(2\delta_i) \sinc (2\pi f_m T)]  \,,
\end{align}
for the entries on the diagonal, and 
\begin{align}
    \Gamma_{\alpha_i \delta_i} 
    & =  - \rho_i^2  F_i A_{mi} \frac{f_i}{f_m}  \cos(\delta_i) \sinc(\pi f_m T)
    \, , \\
    \Gamma_{f_i \delta_i } 
    & =  - \pi T \rho_i^2  F_i A_{mi} \frac{f_i}{f_m}  \sin(\delta_i) \sinc'(\pi f_m T)
    \, , \\
    \Gamma_{\dot{f}_i \delta_i } 
    & = -\pi (\pi T)^2 \rho_i^2 F_i A_{mi}  \frac{f_i}{f_m} \cos(\delta_i) \sinc''(\pi f_m T) 
    \,, \\
    \Gamma_{\alpha_i A_{mi}} & = - \rho_i^2 F_i \frac{f_i}{f_m} \sin (\delta_i) \sinc(\pi f_m T)\,, \\
    \Gamma_{f_i A_{mi}} & = \pi T \rho_i^2 F_i \frac{f_i}{f_m} 
      \cos (\delta_i)\sinc'(\pi f_m T) \, , \\
    \Gamma_{\dot{f}_i A_{mi} } 
    & = - \pi (\pi T)^2 \rho_i^2 F_i \frac{f_i}{f_m} \sin(\delta_i) \sinc''(\pi f_m T) 
    \, ,
\end{align}
for the off-diagonal entries in each block.
The remaining terms are zero to leading order.

Computing these entries requires the application of the various approximations stated in the text, in particular that $f_i T \gg 1$ in all cases, resulting in narrow carrier wave peaks, all well separated from each other.
For example, we have
\begin{align}
    \Gamma_{A_{i}A_{j}} & = \langle \partial_{A_i} \tilde h | \partial_{A_j} \tilde h \rangle 
    = \frac{\langle \tilde h_i | \tilde h_j \rangle }{A_i A_j} 
    \approx \frac{\delta_{ij} \langle \tilde h_i | \tilde h_i \rangle}{A_i^2}  + O(A_m) 
     \notag \\ &
    = \delta_{ij}
    \frac{\rho_i^2}{A_i^2} + O(A_m) \,. 
\end{align}
For those terms which involve derivatives of the finite-time delta functions, the following identities are useful:
\begin{align}
   \int_{-\infty}^{\infty} \frac{\dd \sinc(x+a)}{\dd x} \sinc (x+b)  \dd x  =&  \pi \sinc'x|_{x = a-b}\,, 
    \\ 
     \int_{-\infty}^{\infty} \frac{\dd \sinc(x+a)}{\dd x} \frac{\dd \sinc (x+b)}{\dd x} \dd x  = &- \pi \sinc''x|_{x = a-b}\,.
\end{align}
These can be derived by differentiating Eq.~\eqref{eq:SincIdentity} under the integral.

\subsection{Fast evolution of the modulating wave}

In the case where we neglect the carrier term of the modulation (the pulsar term in the pulsar timing array literature), the parameters $\gamma_i$ which differentiate the $A_{mi}$ and add an unmeasurable phase term to the $\delta_i$ are absent. 
While the amplitude of the sideband varies from carrier to carrier, this variation depends only on the $F_i$, which can be determined from the sky location and which we therefore neglect from our analysis.
Meanwhile, the phase of the modulation encoded in $\delta_i$ depends only on a common phase $\alpha_m$ and the phases $\beta_i$, which depend on sky location and polarization content of the modulating wave.
Thus for our analysis, the sidebands of all the carriers share a common unknown amplitude parameter $A_m$ and phase term $\alpha_m$.
Our parameter set is $\theta^a = \{ A_1, \alpha_1, f_1, \dot f_1, A_2, \dots, A_m, \alpha_m \}$.

The Fisher matrix for our total signal $\tilde h$ breaks into a $4N \times 4N$ block describing the Galactic binary parameters, a rectangular $4N \times 2$ matrix mixing the binary parameters with those of the modulating wave, and a final $2 \times 2$ block with the modulating wave parameters.
We need to keep terms only to leading order in $A_m$.
The $4N \times 4N$ block is block diagonal with a $4 \times 4$ sub-block for each carrier. 
These sub-blocks have diagonal entries
\begin{align}
	\Gamma_{A_i A_i} & = \frac{\rho_i^2}{A_i^2} \,, \\
	\Gamma_{\alpha_i \alpha_i} & = \rho_i^2\,, \\
	 \Gamma_{f_i f_i} & 
    =\frac{(\pi T)^2 \rho_i^2 }{3} \,, \\
    \Gamma_{\dot{f}_i \dot{f}_i}  &
     = \frac{\pi^2}{5} (\pi T )^4  \rho_i^2  \, ,
\end{align}
and non-vanishing off-diagonal entries
\begin{align}
    \Gamma_{\alpha_i \dot{f}_i} 
    & = -\frac{\pi}{3} (\pi T)^2 \rho_i^2 
    \, .
\end{align}
The off-diagonal terms that couple the $4N$ binary parameters with $A_m$ and $\alpha_m$ are
\begin{align}
    \Gamma_{\alpha_i A_m} & = - \rho_i^2 F_i \frac{f_i}{2 f_m} \sin (\delta_i) \sinc(\pi f_m T)\,, \\
    \Gamma_{f_i A_m} & = \pi T \rho_i^2 F_i \frac{f_i}{2 f_m} 
      \cos (\delta_i)\sinc'(\pi f_m T) \, , \\
    \Gamma_{\dot{f}_i A_m } 
    & = - \pi (\pi T)^2 \rho_i^2 F_i \frac{f_i}{2f_m} \sin(\delta_i) \sinc''(\pi f_m T) \, , \\
    \Gamma_{\alpha_i \alpha_m} 
    & =  - \rho_i^2  F_i A_m \frac{f_i}{2f_m}  \cos(\delta_i) \sinc(\pi f_m T)
    \, , \\
    \Gamma_{f_i \alpha_m } 
    & =  - \pi T \rho_i^2  F_i A_m \frac{f_i}{2f_m}  \sin(\delta_i) \sinc'(\pi f_m T)
    \, , \\
    \Gamma_{\dot{f}_i \alpha_m } 
    & = -\pi (\pi T)^2 \rho_i^2 F_i A_m  \frac{f_i}{2f_m} \cos(\delta_i) \sinc''(\pi f_m T)
    \, .
\end{align}
Finally, the terms that involve only the modulating source are
\begin{align}
    \Gamma_{A_m A_m} & = 
    \sum_i 
    \frac{\rho_i^2 F_i^2 f_i^2}
    {8 f_m^2} 
    [1 -\cos(2\delta_i) \sinc (2\pi f_m T)]   \,, \\
    \Gamma_{A_m \alpha_m} &
    = \sum_i \frac{A_m \rho_i^2  F_i^2 f_i^2}{8 f_m^2} \sin(2 \delta_i) \sinc(2 \pi f_m T) \,,
    \\
    \Gamma_{\alpha_m \alpha_m} & =  \sum_i \frac{A_m^2 \rho_i^2  F_i^2 f_i^2}{8 f_m^2}[1 + \cos(2\delta_i)  \sinc(2 \pi f_m T)] \,.
\end{align}

\section{Inverting the Fisher matrix}
\label{sec:InverseAppx}

In Appendix~\ref{sec:FisherApp} we provide the Fisher matrix entries for the measurement of the gravitational wave timing array signal $\tilde h$.
There we treat two extreme cases, the fast and slow cases for the evolution of the modulating wave. 
Here we discuss the aspects of inverting these matrices needed for our sensitivity estimates.

In the slow case, our goal is to compute the entries $\Sigma_{A_{mi}A_{mi}}$ of the covariance matrix $\Sigma_{ab} = (\Gamma^{-1})_{ab}$.
These entries are the squared measurement errors $\sigma_{A_{mi}}$ of the amplitudes $A_{mi}$. 
Since the Fisher matrix is block-diagonal in this case, inversion can be carried out blockwise, and it is straightforward to get the entries $\Sigma_{A_{mi}A_{mi}}$.
In the fast case, the common parameters $\alpha_m$ and $A_m$ couple together the $N$ blocks which correspond to each Galactic binary, and the inversion of $\Gamma_{ab}$ to get $\Sigma_{A_m A_m}$ is more involved.

In the fast case, the Fisher matrix breaks into pieces as follows.
The parameters describing the individual binaries form a $4N \times 4N$ block $\boldsymbol a$ at the upper-left of ${\boldsymbol \Gamma}$, and this sub-matrix is itself block-diagonal, since the individual binaries do not correlate with each other in our approximation.
Denote each of the $N$ $4\times4$ blocks as ${\boldsymbol a}_i$, with $i$ indexing the Galactic binaries.
Next, the last $M$ columns of the first $4N$ rows of ${\boldsymbol \Gamma}$ form a $4N \times M$ matrix ${\boldsymbol b}$, which couples the binaries into the parameters of the modulating wave.
This array itself breaks into $N$ $4 \times M$ blocks ${\boldsymbol b}_i$, with entries such as $(b_1)_{A_1 A_m} = \Gamma_{A_1 A_m}$.
Since ${\boldsymbol \Gamma}$ is symmetric, the first $4N$ columns of the final $M$ rows are ${\boldsymbol b}^\top$, made up of $N$ arrays ${\boldsymbol b}^\top_i$.
Finally, in the lower right we have an $M \times M$ matrix ${\boldsymbol c}$ which covers only the modulating wave parameters.

Now $\Sigma_{A_m A_m}$ sits in the lower-right $M\times M$ block of ${\boldsymbol \Sigma} = {\boldsymbol \Gamma}^{-1}$. Denote these $M\times M$ entries of ${\boldsymbol \Sigma}$ as ${\boldsymbol s}$.
A standard matrix identity, when applied to our decomposition of ${\boldsymbol \Gamma}$, yields
\begin{align}
	{\boldsymbol s} & = \boldsymbol d^{-1} \,, \\
	{\boldsymbol d} &= {\boldsymbol c} - {\boldsymbol b}^\top {\boldsymbol a}^{-1}{\boldsymbol b} 
	=  {\boldsymbol c} - \sum_{i=1}^N {\boldsymbol b}_i^\top {\boldsymbol a}_i^{-1}{\boldsymbol b}_i \,,
\end{align}
where we have defined a useful auxiliary matrix ${\boldsymbol d}$.
The decomposition of the inverse into sums over the contribution from each Galactic binary makes the inversion of the Fisher matrix straightforward: the Fisher matrix can be built using a single Galactic binary and inverted, and in the final solution we need only to sum over the binary indices.

For example, consider a Fisher matrix where we include two parameters of the modulating wave, $A_m$ and $\alpha_m$, and for brevity remove $\dot f_i$ from our parameter list.
Then, recalling that $\Gamma_{A_i A_m} = 0 = \Gamma_{A_i \alpha_m}$ at leading order, the covariance $\Sigma_{A_m A_m}$ is given by
\begin{align}
\Sigma_{A_m A_m} & = \left[ d_{A_m A_m} - \frac{d_{A_m\alpha_m}^2}{d_{\alpha_m\alpha_m}} \right]^{-1}
\,, \\
d_{A_m A_m}
	& = \Gamma_{A_m A_m} - 
		\sum_i 
		\left( 
		\frac{\Gamma_{A_m \alpha_i}^2} {\Gamma_{\alpha_i \alpha_i}}
		+\frac{\Gamma_{A_m f_i}^2}{\Gamma_{f_i f_i}}
		\right) 
		\,,
\end{align}
\begin{widetext}
\begin{align}
\frac{d_{A_m\alpha_m}^2}{d_{\alpha_m\alpha_m}} & = 
		\left[\Gamma_{\alpha_m \alpha_m} 
		- \sum_i
		\left( 
		\frac{\Gamma_{\alpha_m \alpha_i}^2}{\Gamma_{\alpha_i \alpha_i}} 
		+ \frac{\Gamma_{\alpha_m f_i}^2}{\Gamma_{f_i f_i}} 
		\right) 
		\right]^{-1} 
		\left[ \Gamma_{A_m\alpha_m} - \sum_i \left( \frac{\Gamma_{A_m \alpha_i} \Gamma_{\alpha_m \alpha_i}}{\Gamma_{\alpha_i \alpha_i}}
		+ \frac{\Gamma_{A_m f_i} \Gamma_{\alpha_m f_i}}{\Gamma_{f_i f_i}}\right) \right]^2 \,.
	\end{align}
Similar, but more involved expressions give the case we treat in the text, where we include $\dot f_i$, but the main point remains: the inversion can be carried out as if for a single Galactic binary and the common parameters, and then summing any term involving the binary parameters over all the binaries in the network.
This approach gives us the full expressions for our measurement uncertainties given below.

\section{Full expression for the variance of the modulating amplitude}
\label{sec:FullSigmaAppx}

In order to present the full expressions for the covariance matrix entry $\Sigma_{A_m A_m}$ (in the case of a source of modulating waves with fast evolution) and the variance $\sigma_{A_{mi}}$ (in the case of a source of modulating waves with slow evolution), it is useful to define an auxiliary function,
\begin{align}
\label{eq:SensitivityShape}
    g(x_m, \delta_i)
    & = \Bigg[  \Bigg(x_m^6-6 x_m^4-15 x_m^2+\left(6 x_m^4-75 x_m^2+45\right) \cos (2 x_m) +\left(x_m^4-60 x_m^2+180\right) x_m \sin (x_m) \cos (x_m)-45\Bigg)  \nonumber \\
    & \quad \times \Bigg(-x_m \left(x_m^2-12\right) \sin (2 x_m)-6 \left(x_m^2-1\right) \cos (2 x_m) +2 \left(x_m^4-3 x_m^2-3\right)\Bigg) \Bigg] \nonumber \\
    &\times \Bigg[ 2 x_m^6 \Bigg(2 \cos (2 \delta_i) \bigg(\left(x_m^2-3\right) \sin (x_m)+3 x_m \cos (x_m)\bigg) \bigg(3 \left(5-2 x_m^2\right) \sin (x_m)+x_m \left(x_m^2-15\right) \cos (x_m)\bigg) \nonumber \\
    &\quad +\left(x_m^2-6\right) \left(2 x_m^2+3\right) x_m^2+6 \left(15-4 x_m^2\right) x_m \sin (2 x_m) +3 \left(x_m^4-24 x_m^2+15\right) \cos (2 x_m)-45\Bigg) \Bigg]^{-1}
\end{align}
with $x_m = \pi f_m T$.

Then in the case of a source with fast evolution, our estimate for the measurement uncertainty comes from $\Sigma_{A_m A_m}$, where we marginalize over \{$A_i,\alpha_i,f_i,\dot f_i,\alpha_m$\} with a $(4N+2)\times(4N+2)$ Fisher matrix. 
It is given by Eq.~\eqref{eq:fullsigmaAm}.
In the case where the source of modulating waves evolves slowly, we need the uncertainty on each $A_{mi}$ measurement, so we compute the entry $\Sigma_{A_{mi} A_{mi}}$ arising from a $6\times6$ Fisher matrix, where we marginalize over $\{A_i,\alpha_i,f_i,\dot f_i,\delta_i\}$, for each binary.
The result is
\begin{align}
 \label{eq:SlowSigmaAm}
  {\sigma_{A_{mi}}^2} ={\Sigma_{A_{mi}A_{mi}}} 
  & = \left[\rho_i^2 F_i^2 (\pi f_i T)^2 g(x_m,\delta_i) \right]^{-1} \,.
\end{align}

\end{widetext}

\bibliographystyle{utphys}
\bibliography{refs}

\end{document}